\documentclass[reprint,nofootinbib,amsmath,amssymb,aps,onecolumn,superscriptaddress,prd]{revtex4-2}

\usepackage{color}
\usepackage{graphicx}
\usepackage{stmaryrd}
\SetSymbolFont{stmry}{bold}{U}{stmry}{m}{n}
\usepackage{comment}
\usepackage{enumitem}
\usepackage{appendix}
\usepackage{multirow}
\usepackage{natbib}
\usepackage{array}

\usepackage{dcolumn}
\usepackage{bm}

\usepackage[T1,T2A]{fontenc}

\DeclareTextSymbol{\textcyrYu}{T2A}{222}
\newcommand{\cYu}{\text\textcyrYu}
\DeclareTextSymbol{\textcyrYa}{T2A}{223}
\newcommand{\cYa}{\text\textcyrYa}
\DeclareTextSymbol{\textcyrD}{T2A}{196}

\DeclareTextSymbol{\textcyrZh}{T2A}{198}
\newcommand{\cZh}{\text\textcyrZh}

\newcommand{\cmnt}[1]{}

\usepackage{color}
\usepackage{graphicx}
\usepackage{stmaryrd}
\usepackage{comment}

\usepackage{dcolumn}
\usepackage{bm}
\usepackage{xcolor}

\begin{document}
\title{Corrections to Hawking radiation from asteroid-mass primordial black holes: analytic and numerical evaluation of the stochastic charge effect}
\date{\today}

\author{Gabriel Vasquez}
\email{gvasquez0223@gmail.com}
\affiliation{
 Center for Cosmology and Astroparticle Physics, The Ohio State University,
 191 West Woodruff Avenue, Columbus OH, 43210, USA
}
\affiliation{
 Department of Physics, The Ohio State University,
 191 West Woodruff Avenue, Columbus OH, 43210, USA
}

\author{Bowen Chen }
\thanks{Co-second authors.}
\affiliation{
 Center for Cosmology and Astroparticle Physics, The Ohio State University,
 191 West Woodruff Avenue, Columbus OH, 43210, USA
}
\affiliation{
 Department of Physics, The Ohio State University,
 191 West Woodruff Avenue, Columbus OH, 43210, USA
}

\author{Cara Nel }
\thanks{Co-second authors.}
\affiliation{
 Center for Cosmology and Astroparticle Physics, The Ohio State University,
 191 West Woodruff Avenue, Columbus OH, 43210, USA
}
\affiliation{
 Department of Physics, The Ohio State University,
 191 West Woodruff Avenue, Columbus OH, 43210, USA
}

\author{Emily Koivu}
\affiliation{
 Center for Cosmology and Astroparticle Physics, The Ohio State University,
 191 West Woodruff Avenue, Columbus OH, 43210, USA
}
\affiliation{
 Department of Physics, The Ohio State University,
 191 West Woodruff Avenue, Columbus OH, 43210, USA
}

\author{Makana Silva}
\affiliation{
 Computational Physics and Methods Group (CAI-2), Los Alamos National Laboratory, Los Alamos NM, 87544, USA
}
\affiliation{%
Center for Theoretical Astrophysics, Los Alamos National Laboratory, Los Alamos, New Mexico 87544, USA}

\author{Christopher M. Hirata}
\affiliation{
 Center for Cosmology and Astroparticle Physics, The Ohio State University,
 191 West Woodruff Avenue, Columbus OH, 43210, USA
}
\affiliation{
 Department of Physics, The Ohio State University,
 191 West Woodruff Avenue, Columbus OH, 43210, USA
}
\affiliation{
 Department of Astronomy, The Ohio State University,
 140 West 18th Avenue, Columbus OH, 43210, USA
}

\begin{abstract}
Hawking radiation sets stringent constraints on Primordial Black Holes (PBHs) as a dark matter candidate in the $M\sim 10^{16}$ g regime based on the evaporation products produced by the black hole, motivating the need to rigorously model the photon, electron, and positron emission spectra. This manuscript is the second in a series of two papers (see Vasquez et~al. [Phys. Rev. D, 112:063002 (2025)] for the first paper in the series) with the goal of proving the stochastic emission of electrons and positrons, known as the "stochastic charge effect" and first predicted by Page [Phys. Rev. D, 16:2402 (1977)] using semi-classical arguments, arises from quantum electrodynamics (QED) on a Schwarzschild spacetime. We derive the corrections to the $e^\pm$ from the relevant term in the Hamiltonian (the long-range monopole: $H_{\rm int,0L}$) to first order in the fine structure constant ($\alpha$), and show that the semi-classical terms are also present in the QED calculation. We also highlight which terms in our analysis only appear within our quantum mechanical approach and cannot be explained otherwise. We find that due to some cancellation of terms, over the range of $2.2\times 10^{16}$--$1.7\times 10^{17}$\,g, the net correction to the $e^\pm$ emission rate is less than 2\%, versus the suppression of up to 5\% previously found by semi-classical arguments.
\end{abstract}

\maketitle

\section{Introduction}

This paper is the second in a series of two papers (see \citet{2024arXiv240709724V}) with the goal of proving that the stochastic emission of charged particles, commonly known as the \textit{stochastic charge effect}, is a consequence of interacting quantum field theory (QFT) on curved spacetime. If a black hole is initially charged or acquires a net charge from the emission of a charged particle (as measured by an external observer using Gauss' law \cite{PhysRevD.8.3259}), then the black hole will acquire an electric potential at the horizon that will perturb the emission rate of subsequent charged particles, where the probability of emitting a charged particle with charge of the same sign as the black hole is more likely,  driving the black hole towards neutrality. This effect was first studied by \citet{PhysRevD.16.2402} in the context of primordial black holes (PBHs) with masses in the range of $M\sim 10^{16}$ g. The Hawking temperature for a Schwarzschild black hole is $T_{H} = 1/(8\pi M) \approx 1\,(10^{16}\,{\rm g}/M_{\rm PBH})\,{\rm MeV}$, so in the $10^{16}$--$10^{17}$\,g mass range, the stochastic charge effect is sourced by the emission of electrons and positrons emanating from the black hole. Our ultimate goal is to derive the $\mathcal{O}(\alpha)$ correction to the electron and positron emission spectra starting from the full quantum electrodynamic (QED) Lagrangian on a Schwarzschild spacetime, where $\alpha \approx 1/137$ is the fine structure constant.
We also want to understand how (and whether) the terms ascribed to the stochastic charge effect  by \citet{PhysRevD.16.2402} arise in QFT.

PBHs have gained a resurgence of interest within the last decade due to their viability as a potential dark matter candidate. They are solutions to standard general relativity, do not require the introduction of a new long-lived elementary particle, and would act like cold dark matter from a cosmological structure formation point of view. They do require a novel formation mechanism, since the ``standard'' inflationary power spectrum cannot produce significant numbers of PBHs --- rather, a break in the inflationary power spectrum \cite[e.g.][]{1975ApJ...201....1C, 1997A&A...318..673Y}, an early matter-dominated era \cite[e.g.][]{1980PhLB...97..383K, 2016ApJ...833...61H}, and/or phase transitions \cite[e.g.][]{1999PhRvD..59l4014J} would be needed. While PBHs may be theoretically formed at any mass (see \citet{carr2026primordialblackholesconstraints} for a summary of the most up-to-date constraints), the asteroid-mass regime ($10^{16}-10^{22}$ g) has gained renewed interest due to the relaxation of previous constraints, allowing PBHs to compose all of the dark matter content \cite{Katz_2018, Montero_Camacho_2019, Sugiyama_2020}. However, recent photometric observations of ultra-faint dwarf galaxies from the Hubble Space Telescope have been used to exclude PBHs in the $10^{18}-10^{20}$ g regime from composing all of the dark matter content in our Universe \cite{Esser_2025}. Additionally, PBHs must survive to the present day in order to be a viable dark matter content, presenting a stringent lower bound where $M \gtrsim 5 \times 10^{14}$ g. This leaves the region of a few$\times 10^{16}$--$10^{18}$ g where PBHs may compose the total dark matter content. 

Constraints in the $\lesssim 10^{17}$\,g window are usually placed by investigating how the emission products from Hawking radiation alter astrophysical observables. For example, gamma ray photons produced from evaporating PBHs may contribute towards the extragalactic background \cite{Carr_2010} while other authors have considered how the local sky brightness is modified \cite{Laha_2020}. Positrons emitted from Hawking radiation would contribute towards the 511 keV annihilation line measured from the Galactic Center, which has been used to set constraints on the PBH abundance \cite{Laha_2019} while {\slshape Voyager 1} can set stringent constraints on $M<10^{16}$ g black holes due to its sensitivity towards low-energy cosmic rays and its location in interstellar space \cite{Boudaud_2019}. The injection of Hawking emitted particles can also modify the ionization history of our Universe which would have significant implications towards our understanding of cosmology \cite{Yang_2022, koivu2025effectsprimordialblackholes}.
Current and future MeV telescopes, such as upcoming COSI mission \cite{2024icrc.confE.745T} and the proposed AMEGO observatory, will have the sensitivity to detect Hawking radiation if enough of the dark matter is in PBHs near the lower end of this window \cite{PhysRevLett.126.171101, 2023JCAP...02..006C}. If a detection is made, it will not only have enormous implications towards unmasking the identity of dark matter and providing concrete evidence that PBHs exist, but would also be the first observational evidence of the existence of Hawking radiation. 

Traditionally, PBH emission spectra are modeled as a thermal ``greybody,'' a blackbody with an additional term that quantifies the energy-dependent cross section for an emitted particle to reflect and be absorbed by the black hole. These types of calculations are streamlined by codes such as {\sc{BlackHawk}} which are used to calculate the primary emission spectra (produced directly from the black hole) and the secondary emission spectra (produced via decays and final state radiation) in a wide range of astrophysical scenarios \cite{Arbey_2019,Arbey_2021, Arbey_2021_part2}. These calculations are performed in the \textit{semi-classical perspective}, where particle formation due to Hawking radiation occurs at the horizon and the likelihood of an emitted particle escaping to $\infty$ is determined by the transmission probability through the effective potential in the radial wave equation.

Despite the popularity of the semi-classical approach, there are paradoxes if it is extended all the way to the horizon \cite{Giddings_2016, giddings1994quantummechanicsblackholes, Mathur_2009,braunstein2009entangledblackholesciphers, Almheiri_2013}. Some authors have suggested that Hawking radiation is sourced by the near-horizon quantum region, the quantum ``atmosphere'' surrounding the black hole \cite{Giddings_2016, Hod_2016, Dey_2017}.
Even in the ``standard'' picture of QFT on curved spacetime (which we use in this paper), it does not make sense to talk about the charge of the horizon changing as a function of time because in any finite time as seen by an external observer no charge could enter or leave the horizon. This raises two questions: in QFT on a Schwarzschild spacetime, with fixed charge on the horizon itself, is it possible to replicate the semi-classical correction to the $e^\pm$ emission spectra, as first calculated by \citet{PhysRevD.16.2402}, entirely from particle interactions outside the horizon? If so, then how does this $\mathcal{O}(\alpha)$ correction to the $e^\pm$ spectrum fit in with the other contributions?



In our previous paper, \citet{2024arXiv240709724V}, we proposed two conjectures on the relation between the semi-classical calculation \cite{PhysRevD.16.2402} and ``full'' QED correction that address these questions. We first promoted the black hole's charge $Z$ to an operator ${\cal \hat Z}$, which contains a contribution from both the charge of the black hole and the surrounding electron-positron plasma, weighted by $2M/r$: charges count toward ${\cal \hat Z}$ if they are very near the horizon, do not count if far away, but their weight varies smoothly in between. The $e^\pm$ contribution is expressed in terms of the familiar fermion creation and annihilation operators.
{\em Conjecture \#1} holds that the semi-classical variance $\langle Z^2 \rangle$ agrees with the variance predicted by the quantum mechanical operator $\langle {\cal \hat Z}^2 \rangle$ in QED. {\em Conjecture \#2} holds that the semi-classical corrections to the emitted $e^\pm$ spectrum from \citet{PhysRevD.16.2402} also appear as contributions in the full QED calculation. \citet{2024arXiv240709724V} concluded with a proof of Conjecture \#1.


The goal of this manuscript is to prove Conjecture $\#$2. We aim to answer the following questions: What is the emission spectra one can derive from the QED formalism on a Schwarzschild spacetime? Are there terms in the emission spectra that agree and can explain the semi-classical prediction, which are they, and what terms cannot be explained semi-classically? When computed numerically, are there significant deviations from the semi-classical perspective? If Conjecture $\#$2 is proven true, and we able to replicate the semi-classical emission spectra first derived in \citet{PhysRevD.16.2402}, we can conclude the origin of stochastic charge phenomenon is understood in the context of quantum field theory on curved spacetime rather than from semi-classical arguments.

\section{Outline and formalism} 


We adopt the same conventions used in \citet{PhysRevD.107.045004}, and in Sec. III of our first paper \cite{2024arXiv240709724V}. We also use the generalized Coulomb gauge, following those papers. Many of the same mathematical expressions used in our previous manuscript to prove Conjecture $\#$1 are used here. We provide a table (Tab. \ref{tab:math_terms_from_first_paper}) listing relevant terms the reader should know, a brief description of their physical meaning and location in \cite{2024arXiv240709724V}, and relevant sections where an interested reader can learn more of how these terms are derived and their physical meaning.

For the photon spectrum \cite{PhysRevD.107.045004}, only the electrodynamic sector ($A_i$ in Coulomb gauge) entered at $\mathcal{O}(\alpha)$, while the electrostatic sector ($\Phi$ in Coulomb gauge) of the QED Lagrangian does not contribute. In contrast, for the $e^\pm$, there is a coupling between the electron field and the $\Phi$ sector of the QED Lagrangian which would produce an $\mathcal{O}(\alpha)$ correction. Each sector can be mediated by a (possibly virtual) photon of angular momentum $\ell$, and the $\ell=0$ part of the $\Phi$ sector is special in that we split it into a long-range and short-range component. The stochastic charge effect arises from the $\Phi$ sector, $\ell=0$, long-range (interaction Hamiltonian $H_{\rm int, 0L+bdy}$).

Our paper is structured in the following way:

\begin{enumerate}[label=\Roman*., start=3]
    \item We derive the perturbed $e^\pm$ emission spectra due to the stochastic charge effect starting from Eq.(61) of our first paper \cite{2024arXiv240709724V} and categorize which contributions are from free-field terms and which are not. We then express these terms, when relevant, in the context of first and second derivatives of the transmission probability ${\mathbb T}_{\frac12 kh}$ and unperturbed electron phase space densities $f_{X}(h)$ (which is only nonzero when $X=\rm up$).
    \item We discuss the numerical methods used towards calculating our derived emission spectra.
    \item We provide a discussion on our derived spectra. We discuss the physical interpretation of terms we derived analytically and our numerical results. 

    \item We conclude our discussion proving Conjecture $\#$2.
\end{enumerate}

\begin{table}[ht]
\centering
\begin{tabular}{|p{2.3cm}|p{8.7cm}|p{6.0cm}|}
 \multicolumn{3}{c}{\bfseries Important Mathematical Terms from \citet{2024arXiv240709724V}} \\
 \hline
 Symbol & Meaning & Reference \\
 \hline
 \multicolumn3c{\bfseries Quantum numbers} \\
 \hline
 $m$ & Fermion $z$-angular momentum. & Standard quantum mechanics notation \\
 \hline
 $k$ & Fermion total angular momentum: $|k|=j+\frac12$, $k>0$ if electron spin parallel to total angular momentum and $k<0$ if anti-parallel. $k$ is an integer. & \citet{1957RvMP...29..465B}, Eqs. (35, 37) \\
 \hline
 $h$ & Electron energy (as measured at $\infty$). & \citet{1957RvMP...29..465B}, Eq. (32) \\
 \hline
 \multicolumn3c{\bfseries Infinitesimals and limiting functionals} \\
 \hline
 $\epsilon$ & Small imaginary component ($h-h' \approx \pm i \epsilon$) introduced due to the singularities induced by the Coulomb interaction not vanishing at the horizon.  & \citet{2024arXiv240709724V}, \S VIE \\
 \hline
 $\eta$ & Small imaginary component ($h-h' \approx i\eta$) introduced due to singularities caused by utilizing the steady-state approximation. & \citet{2024arXiv240709724V}, \S VIE \\
\hline
$\delta_\epsilon(h-h')$ & Lorentzian (Dirac delta) function with width $\epsilon$ ($\epsilon \rightarrow$ 0). & \citet{2024arXiv240709724V}, Eq. (C11)
\newline\\
 \hline
 \multicolumn3c{\bfseries Operators} \\
 \hline
 $\hat {\cal Z}$ & Charge of the black hole (${\cal Q}_-)$ plus its surroundings weighted by $2M/r$: $-e\hat{\cal Z} = {\cal Q}_- + \int (2M/r) J^t \sqrt{-g}\,d^3x$. & \citet{2024arXiv240709724V}, Eq. (42) \\
 \hline
 \multicolumn3c{\bfseries Mode integrals} \\
 \hline
${\cal I}^{(1)}_{XX'k}(h,h')$ & Radial overlap integral between different fermionic wave modes with positive energies (weighted by $2M/r$), relevant for counting +1 or -1 for a given electron mode in $\cal{\hat Z}$. & \citet{2024arXiv240709724V}, Eq. (47)  \\
\hline
${\cal I}^{(2)}_{XX'k}(h,h')$ &  Radial overlap integral between positive-energy and negative-energy electron modes (in the Dirac sea), relevant for depicting pair-production and pair-annihilation (weighted by $2M/r$). & \citet{2024arXiv240709724V}, Eq. (47)  \\
\hline
${\cal A}_{XX'k}(h,h')$ & ${\cal I}^{(1)}_{XXk}(h,h')$ multiplied by $h-h'$ to remove the pole (divergence due to integrating to the horizon at $r_\star=-\infty$). & This paper, Eq.~(\ref{eq:A-def}) \\
\hline
$\Lambda_{XX'k}(h,h')$ & Nonsingular component of ${\cal I}^{(1)}_{XX'k}(h,h')$. Related to, but not identical to ${\cal A}_{XX'k}(h,h')$. & This paper, Eq.~(\ref{eq:A-table}) \\
 \hline
 \multicolumn3c{\bfseries Correlation functions} \\
 \hline
 $\langle \hat {\cal Z}^2 \rangle$ & Variance of charge operator ${\cal \hat Z}$. & \citet{2024arXiv240709724V}, Eq. (127, 128)
\\
 \hline
 $f_{X}(h)$ & Unperturbed phase space density of fermions: $f_{\rm in}(h) = 0$ and $f_{\rm up}(h) = (e^{8\pi Mh} +1)^{-1}$. &  Also denoted $f_{XX'km}^{e^\pm}(h)$ \cite{2024arXiv240709724V, PhysRevD.107.045004}. This is simplified here since it is only nonzero if $X=X'$ and does not depend on $k$ or $m$. \\
\hline
$ \Gamma^{{\cal \hat Z}b^\dagger b (k)}_{XX'}(h,h')$ & Correlation function of $\langle {\cal \hat Z} {\hat b}_{Xkmh}^\dagger {\hat b}_{X'kmh'} \rangle$ (these correlators are diagonal in $k$ and $m$ and does not depend on $m$ due to spherical symmetry). & \citet{2024arXiv240709724V}, Eq. (84) \\
\hline
$ \Gamma^{{\cal \hat Z}bd (k)}_{XX'}(h,h')$ & Correlation function of $\langle {\cal \hat Z} {\hat b}_{Xkmh} {\hat d}_{X'kmh'} \rangle$. & \citet{2024arXiv240709724V}, Eq. (95) \\
\hline
$ \Gamma^{{\cal \hat Z}d^\dagger b^\dagger (k)}_{XX'}(h,h')$ & Correlation function  $\langle {\cal \hat Z} {\hat d}_{Xkmh}^\dagger {\hat b}^\dagger_{X'kmh'} \rangle$. & \citet{2024arXiv240709724V}, Eq. (102) \\
 \hline
\end{tabular}
\caption{Important mathematical expressions used in \citet{2024arXiv240709724V} and needed here. We provide a brief summary of the physical interpretation of these terms while also providing a reference to the full expression and details. 
}
\label{tab:math_terms_from_first_paper}
\end{table}

\section{The evolution equation}
\label{sec:evol}

The goal of this section is to calculate the $\mathcal{O}(\alpha)$ perturbation to the $e^\pm$ emission spectrum due to the long-range monopole potential Hamiltonian $H_{\rm int, 0L+bdy}$. The Hamiltonian is given in terms of the effective charge operator ${\cal Z}$ by Eq.~(40) of \citet{2024arXiv240709724V}:
\begin{equation}
H_{\rm int, 0L+bdy} = \frac{e^2}{16\pi M} \hat{\mathcal Z}^2.
\end{equation}
The charge operator may be expressed using Eqs.~(41, 43) or (46) of \citet{2024arXiv240709724V}:
\begin{eqnarray}
{\mathcal Z} &=& - \int \frac{2M}r :\!\psi^\dagger(x) \bar\psi(x)\!: r^2\sqrt{1-\frac{2M}r}\sin\theta\,dr_\star \,d\theta\,d\phi  - \frac{{\cal Q}_-}e
\nonumber \\
&=& - \int \frac{dh\,dh'}{(2\pi)^2} \sum_{XX'km}
\Bigl[
{\cal I}^{(1)}_{XX'k}(h,h')  \hat b^\dagger_{Xkmh} \hat b_{X'kmh'} - {\cal I}^{(1)}_{XX'-k}(h,h') \hat d^\dagger_{Xkmh} \hat d_{X'kmh'}
\nonumber \\
&& ~~~~+ {\cal I}^{(2)}_{XX'k}(h,h')  \hat b^\dagger_{Xkmh} \hat d^\dagger_{X'-kmh'} + {\cal I}^{(2)}_{XX'-k}(h,h')
\hat d_{Xkmh} \hat b_{X'-kmh'}
\Bigr]
 - \frac{{\cal Q}_-}e;
\end{eqnarray}
here ${\cal Q}_-$ is the (unchanging) charge on the horizon, and $\hat b$ and $\hat d$ are fermion annihilation operators. The radial coupling integrals are given by Eq.~(47) of \citet{2024arXiv240709724V}
\begin{equation}
\begin{aligned}
& \mathcal{I}_{X X^{\prime} k}^{(1)}\left(h, h^{\prime}\right) \equiv \frac{1}{\sqrt{4 h h^{\prime}}} \int_{-\infty}^{\infty} d r_{\star} \frac{2 M}{r}\left(F_{X k h}^* F_{X^{\prime} k h^{\prime}}+G_{X k h}^* G_{X^{\prime} k h^{\prime}}\right) \quad \text { and } \\
& \mathcal{I}_{X X^{\prime} k}^{(2)}\left(h, h^{\prime}\right) \equiv \frac{1}{\sqrt{4 h h^{\prime}}} \int_{-\infty}^{\infty} d r_{\star} \frac{2 M}{r}\left(F_{X k h}^* G_{X^{\prime}-k h^{\prime}}^*+G_{X k h}^* F_{X^{\prime}-k h^{\prime}}^*\right),
\end{aligned}
\label{eq:I12-expression}
\end{equation}
where $(F, G)$ represent the radial solutions of the Dirac equation.

We begin our calculation starting from the $\mathcal{O}(\alpha)$ correction to the outgoing electron emission spectrum given by Eq.~(61) of \citet{2024arXiv240709724V}:
\begin{equation}
\label{eq:electron_spectra_first_eq}
\left.\frac{dN_-}{dh\,dt}\right|_{\rm 0L}
= - \frac{\alpha}{2\pi M}
\sum_k (2j+1) \Im \int_0^\infty \frac{dh'}{2\pi}
\sum_{XX'Y} w_{Xkh}w_{Ykh}^\ast
\left[{\cal I}^{(1)}_{XX'k}(h,h') W^{(k)}_{YX'}(h,h')
-
{\cal I}^{(2)}_{XX'k}(h,h') P^{(k)}_{YX'}(h,h')\right],
\end{equation}
where $X$, $X'$, and $Y$ range over $\{{\rm in,up}\}$, the coefficients are $w_{{\rm up},kh}=T_{\frac12 kh}$ and $w_{{\rm in},kh}=R_{\frac12 kh}$, and $\alpha = e^2/4\pi$. Here we have defined the Hermitian components of the correlation functions:
\begin{equation}
\begin{aligned}
& W^{(k)}_{YX'}(h,h') = \frac12 \left[ \Gamma^{{\cal Z}b^\dagger b(k)}_{YX'}(h,h') + \Gamma^{{\cal Z}b^\dagger b(k)\ast}_{X'Y}(h',h)\right]
~~~~{\rm and}\\
& P^{(k)}_{YX'}(h,h') = \frac12 \left[ \Gamma^{{\cal Z}d^\dagger b^\dagger(k)}_{X'Y}(h',h) + \Gamma^{{\cal Z}b d(k)\ast}_{YX'}(h,h')\right].
\end{aligned}
\label{eq:W-}
\end{equation}
Solving Eq.~(\ref{eq:electron_spectra_first_eq}) will require us to integrate across a double pole located at $h=h'$ for terms without a copy of $\langle \hat{\cal Z}^2 \rangle$, while terms with a copy of $\langle \hat{\cal Z}^2 \rangle$ will contain a triple pole located at $h=h'$ on the complex plane. Therefore, one must use an expression for ${\cal I}^{(1)}_{XX'k}(h,h')$ that correctly treats the singularity to one higher order in $\epsilon$. The relevant expression is presented in Appendix~\ref{app:E-Lambda} and is
\begin{equation}
    {\cal I}^{(1)}_{XX'k}(h,h') = \frac{{\cal A}_{XX'k}(h,h') -i \epsilon \Lambda_{XX'k}(h,h)}{h-h'-i\epsilon} + 2\pi \delta_\epsilon(h-h') \delta_{X,\rm up} \delta_{X', \rm up}.
\label{eq:I(1)-modified}
\end{equation}
(Recall that $\boldsymbol\Lambda_k(h,h)$ is Hermitian, and see Table~\ref{tab:math_terms_from_first_paper} for how we decompose the singularity structure of ${\cal I}^{(1)}_{XX'k}(h,h')$.)
When Eq.~(\ref{eq:I(1)-modified}) is included into the definition of $\Gamma^{{\cal Z}b^\dagger b(k)}_{XX'}(h,h')$ (Eq.~(84) of \citet{2024arXiv240709724V}), the modified version of the correlation function is given by
\begin{equation}
\begin{split}
    \Gamma^{{\cal Z}b^\dagger b (k)}_{XX'}(h,h') =& - \frac{{\cal A}^\ast_{XX'k}(h,h')}{h-h'+i\eta} \left[1 - f_{X}(h) \right] f_{X'}(h') \\
    \ & - \frac{\alpha \langle {\cal \hat Z}^2 \rangle}{2M} \frac{f_{X}(h) - f_{X'}(h')}{h-h'+i\eta} \left[ \frac{{\cal A}^\ast_{XX'k}(h,h') +i \epsilon \Lambda^\ast_{XX'k}(h,h)}{h-h'+i\epsilon} + 2\pi \delta_\epsilon(h-h') \delta_{X,\rm up} \delta_{X',\rm up} \right].
\end{split}
\end{equation}
(Only the contributions that will have a triple pole need to include the $\epsilon \Lambda$ term.)
Thus, Eq.~(\ref{eq:W-}) becomes
\begin{equation}
\begin{split}
    W^{(k)}_{YX'}(h,h')  =& \frac{{\cal A}^\ast_{YX'k}(h,h')}{h-h'+i\eta} \left[f_Y(h) f_{X'}(h') - \frac{f_Y(h) + f_{X'}(h')}{2} \right] \\
    \ & - \frac{\alpha \langle {\cal \hat Z}^2 \rangle}{2M}\frac{f_{Y}(h) - f_{X'}(h')}{h-h'+i\eta} \left[ \frac{{\cal A}^\ast_{YX'k}(h,h') + i\epsilon \Lambda^\ast_{YX'k}(h,h)}{h-h'+i\epsilon} + 2\pi \delta_\epsilon(h-h') \delta_{Y,\rm up} \delta_{X',\rm up} \right]
\label{eq:W-modified}
\end{split}
\end{equation}
which is the Hermitian component of our modified $\Gamma^{{\cal Z}b^\dagger b(k)}_{XX'}(h,h')$. The next step is to solve for the contribution due to pair-production and pair-annihilation. The ${\cal I}^{(2)}$ terms are non-singular and do not require any special consideration when $h \approx h'$; instead, one has
\begin{equation}
P^{(k)}_{YX'}(h,h') = {\cal I}^{(2)\ast}_{YX'}(h,h')\left[ \frac{1-f_Y(h)-f_{X'}(h')}{2}
+ f_Y(h) f_{X'}(h')
\right]
+ \frac{\alpha\langle{\cal\hat Z}^2\rangle}{ 2M} \frac{f_Y(h) + f_{X'}(h') - 1}{h+h'}\, {\cal I}^{(2)\ast}_{YX'}(h,h')
\end{equation}
which is well-behaved for positive energies.

Therefore, the resulting perturbation to the emitted electron spectrum has two classes of terms: an $\sim\alpha$ contribution resulting from the self-interaction of the electron field via the monopole moment of the long-range Coulomb potential and an $\sim \alpha^2 \langle \hat{\cal Z}^2 \rangle$ correction associated with the stochastic charge effect. The resulting perturbation may be written as
\begin{equation}
\left.\frac{dN_-}{dh\,dt}\right|_{\rm 0L}
= \frac{1}{2\pi}\sum_k (2j+1) \left[ -\frac{\alpha}{M}B_{0}(k,h) + \frac{\alpha^2 \langle\hat{\cal Z}^2\rangle}{2M^2}B_{1}(k,h) \right].
\label{eq:NHT}
\end{equation}
where both terms are $\mathcal{O}(\alpha)$, since $\langle \hat{\cal Z}^2 \rangle \sim \mathcal{O}(\alpha^{-1})$ to lowest order. The expressions for $B_0(k,h)$ and $B_1(k,h)$ are given by
\begin{align}
B_{0}(k,h) =& \Im \int_0^\infty \frac{dh'}{2\pi} \Biggl\{ \sum_{XX'Y}
w_{Xkh}w^\ast_{Ykh} \frac{{\cal A}_{XX'k}(h,h'){\cal A}^\ast_{YX'k}(h,h')}{(h-h'+i\eta)(h-h'-i\epsilon)}
\left[f_Y(h) f_{X'}(h') - \frac{f_Y(h) + f_{X'}(h')}{2} \right]
\nonumber \\
&~~ + \sum_Y T_{\frac12,kh}w^\ast_{Ykh} \frac{2\pi\delta_\epsilon(h-h'){\cal A}_{Y,{\rm up},k}^\ast(h,h')}{h-h'+i\eta}
\left[f_Y(h) f_{\rm up}(h') - \frac{f_Y(h) + f_{\rm up}(h')}{2} \right]
\nonumber \\ &~~
+ \sum_{XX'Y} w_{Xkh} w_{Ykh}^\ast {\cal I}^{(2)}_{XX'k}(h,h') {\cal I}^{(2)\ast}_{YX'k}(h,h')
\left[ \frac{1-f_Y(h)-f_{X'}(h')}{2}
+ f_Y(h) f_{X'}(h')
\right]
\Biggr\}
\label{eq:B0}
\end{align}
and
\begin{align}
B_{1}(k,h) =& \Im \int_0^\infty \frac{dh'}{2\pi} \Biggl\{ \sum_{XX'Y} w_{Xkh} w_{Ykh}^\ast \frac{[f_Y(h)-f_{X'}(h')]{\cal A}_{XX'k}(h,h'){\cal A}^\ast_{YX'k}(h,h')}{(h-h'+i\eta)(h-h'+i\epsilon)(h-h'-i\epsilon)}
\nonumber \\
& -i \epsilon \sum_{XX'Y} w_{Xkh} w_{Ykh}^\ast \frac{[f_Y(h)-f_{X'}(h')]{\Lambda}_{XX'k}(h,h){\cal A}^\ast_{YX'k}(h,h')}{(h-h'+i\eta)(h-h'+i\epsilon)(h-h'-i\epsilon)}
\nonumber \\
& + i\epsilon \sum_{XX'Y} w_{Xkh} w_{Ykh}^\ast \frac{[f_Y(h)-f_{X'}(h')]{\cal A}_{XX'k}(h,h'){\Lambda}^\ast_{YX'k}(h)}{(h-h'+i\eta)(h-h'+i\epsilon)(h-h'-i\epsilon)}
\nonumber \\
& + \sum_Y T_{\frac12,kh} w_{Ykh}^\ast [f_Y(h)-f_{\rm up}(h')]
\frac{2\pi\delta_\epsilon(h-h') {\cal A}_{Y,{\rm up},k}^\ast(h,h')}{(h-h'+i\eta)(h-h'+i\epsilon)}
\nonumber \\
& + i\epsilon\sum_Y T_{\frac12,kh}w_{Ykh}^\ast [f_Y(h)-f_{\rm up}(h')]
\frac{2\pi\delta_\epsilon(h-h') {\Lambda}_{Y,{\rm up},k}^\ast(h)}{(h-h'+i\eta)(h-h'+i\epsilon)}
\nonumber \\
& + \sum_X w_{Xkh}T^\ast_{\frac12,kh} [f_{\rm up}(h)-f_{\rm up}(h')] \frac{2\pi\delta_\epsilon(h-h'){\cal A}_{X,{\rm up},k}(h,h')}{(h-h'+i\eta)(h-h'-i\epsilon)}
\nonumber \\
& + |T_{\frac12,kh}|^2 [f_{\rm up}(h)-f_{\rm up}(h')] \frac{[2\pi\delta_\epsilon(h-h')]^2}{h-h'+i\eta}
\nonumber \\
& + \sum_{XX'Y} w_{Xkh} w_{Ykh}^\ast {\cal I}^{(2)}_{XX'k}(h,h'){\cal I}^{(2)\ast}_{YX'k}(h,h') \frac{f_Y(h)+f_{X'}(h')-1}{h+h'}
\Biggr\}.
\label{eq:B1}
\end{align}
Inspecting the structure of the integrand of each function, $B_0$ contains a double pole across $h'=h+i\eta$ and $h'=h-i\epsilon$ while $B_1$ contains a triple pole across $h'=h+i\eta$, $h'=h+i
\epsilon$, and $h'=h-i\epsilon$. 
\subsection{The singular part}

We can use the expressions in Appendix~\ref{app:singular} to solve for the singular parts of $B_0$ and $B_1$. Specifically, we know that integrals with only $h-h'+i\eta$ and $h-h'+i\epsilon$ in the denominator will be finite, because the integrand can be analytically continued and the path of integration deformed to not pass near the poles (as opposed to integrals with both $h-h'+i\eta$ and $h-h'-i\epsilon$, where the path of integration "threads the needle" between the singularities). As currently written, the pole structure of Eq.~(\ref{eq:B0}) and Eq.~(\ref{eq:B1}) will produce singular terms that may or may not cancel. Therefore, the goal of this calculation is to prove that the only nonzero contribution to $B_0$ and $B_1$ is from the non-singular part of these functions, where the poles are all on the same side of the complex plane.

In general, for any integral expression $expr$, we define the singular part as
\begin{equation}
{\mathbb V}expr = expr - [expr:\,h-h'-i\epsilon \rightarrow h-h'+i\epsilon, \delta_\epsilon(h-h')\rightarrow 0].
\label{eq:V-expr}
\end{equation}
As shown in Appendix~\ref{app:singular}, the types of integrals that appear here have singular contributions of three possible orders: $\epsilon^{-1}\eta^{-1}$, $\eta^{-2}$, and $\eta^{-1}$. Therefore, it is possible to sum up each type of singularity that appears in Eq.~(\ref{eq:B0}) and Eq.~(\ref{eq:B1}). However, it is important to note that ${\mathbb V}expr$ does not have any ${\cal O}(1)$ contributions: all contributions to it are either in one of the three divergent orders, or go to zero in the limit: $\lim_{\eta\rightarrow 0^+}\lim_{\epsilon\rightarrow 0^+}$.

We begin our discussion by focusing on the singularity structure of $B_0(k,h)$. For $B_0$, there is no $\eta^{-2}$ contribution in the integrand, so ${\mathbb V} B_0(k,h)\vert_{\eta^{-2}} = 0$.
However, there is a $ \epsilon^{-1}\eta^{-1}$ and a $ \eta^{-1}$ contribution that both need to be accounted for. To do so, we use Eq.~(\ref{eq:ya1}) and Eq.~(\ref{eq:ya2}) to solve for the $\eta^{-1}$ and $\epsilon^{-1}\eta^{-1}$ contributions, respectively. The result is that all the contributions to the singular part of integral are real, so they do not contribute to $B_0$. Thus
\begin{equation}
{\mathbb V} B_0(k,h) |_{\eta^{-2}} = {\mathbb V}B_0(k,h)|_{\epsilon^{-1}\eta^{-1}} =
{\mathbb V}B_0(k,h)|_{\eta^{-1}} = 0.
\end{equation}
We now turn our attention to the singularity structure of $B_1(k,h)$. We find that the $\eta^{-2}$ term has only contributions from the $1^{\text{st}}$ and $4^{\text{th}}$ lines of Eq.~(\ref{eq:B1}):
\begin{align}
{\mathbb V}B_1(k,h)|_{\eta^{-2}} = & ~\frac12 f_{\rm up}(h) \Im \Bigl[
-i|R_{\frac12,kh}|^2 |{\cal A}_{{\rm in,up},k}(h,h)|^2 
+iR_{\frac12,kh}T^\ast_{\frac12,kh} {\cal A}_{{\rm in,in},k}(h,h) {\cal A}^\ast_{{\rm up,in},k}(h,h) 
\nonumber \\ &~~
- iT_{\frac12,kh}R^\ast_{\frac12,kh} {\cal A}_{{\rm up,up},k}(h,h) {\cal A}_{{\rm in,up},k}^\ast(h,h) 
+ i|T_{\frac12,kh}|^2 |{\cal A}_{{\rm up,in},k}(h,h)|^2 
- T_{\frac12,kh}R^\ast_{\frac12,kh} {\cal A}^\ast_{{\rm in,up},k}(h,h') \Bigr]
\nonumber \\
=& ~\frac12 f_{\rm up}(h) \Im \Bigl[-i|R_{\frac12,kh}|^4|T_{\frac12,kh}|^2 -i |R_{\frac12,kh}|^2|T_{\frac12,kh}|^4 - i|R_{\frac12,kh}|^2 |T_{\frac12,kh}|^4 
\nonumber \\
& ~~+ i |T_{\frac12,kh}|^4|R_{\frac12,kh}|^2 + i |T_{\frac12,kh}|^2|R_{\frac12,kh}|^2\Bigr]
\nonumber \\
=& ~0,
\end{align}
where the final equality results from plugging in the values of Eq.~(\ref{eq:A-val}) and using $|R_{\frac12,kh}|^2 + |T_{\frac12,kh}|^2=1$. For each expression in Eq.~(\ref{eq:B1}), the $\epsilon^{-1}\eta^{-1}$ contribution is the negative of the $\eta^{-2}$ contribution and thus cancels out, except for the $\delta_\epsilon^2$ term. However, since the prefactor is zero at $h'=h$, this term does not contribute due to Eq.~(\ref{eq:DE2}). Therefore,
\begin{equation}
{\mathbb V}B_1(k,h)|_{\epsilon^{-1}\eta^{-1}} = 0.
\end{equation}
Finally, we solve for ${\mathbb V}B_1(k,h)|_{\eta^{-1}}$ which includes {\em derivatives} of ${\cal A}_{XX'k}(h,h')$. Defining the shorthand
\begin{equation}
{\cal E}_{XX'k}(h) =\frac{\partial}{\partial h'}{\cal A}_{XX'k}(h,h') \Bigr|_{h'=h},
\label{eq:E}
\end{equation}
where the explicit form of these derivatives are given by Eq.~(\ref{eq:E2}) and Eq.~(\ref{eq:E-SUB}), and dropping terms that manifestly have no imaginary part, we find
\begin{align}
{\mathbb V}B_1(k,h)\Big|_{\eta^{-1}} =& ~\Im \Bigl\{
\frac12 i f_{\rm up}(h)  |T_{\frac12,kh}|^2 R_{\frac12,kh}T_{\frac12,kh}^\ast {\cal E}^\ast_{{\rm up,in},k}(h)
+\frac12 i f_{\rm up}(h)  |R_{\frac12,kh}|^2|T_{\frac12,kh}|^2 {\cal E}_{{\rm in,in},k}(h)
& [1_{\rm in,up,in}]
\nonumber \\ & ~
+ \frac12i T_{\frac12,kh}R_{\frac12,kh}^\ast|T_{\frac12,kh}|^2 f_{\rm up}(h) {\cal E}^\ast_{{\rm in,up},k}(h,h')- \frac12  i |R_{\frac12,kh}|^2|T_{\frac12,kh}|^2 f_{\rm up}(h) {\cal E}_{{\rm up,up},k}(h,h')
& [1_{\rm up,in,up}]
\nonumber \\ & ~
-\frac 12 i |R_{\frac12 kh}|^2 R_{\frac12 kh}T_{\frac12 kh}^\ast f_{\rm up}(h) \Lambda_{{\rm in,up},k} (h)
& [2_{\rm in,in,up}]
\nonumber \\ & ~
+\frac 12 i |R_{\frac12kh}|^2 |T_{\frac12kh}|^2 f_{\rm up}(h) \Lambda_{{\rm in,in},k}(h) 
& [2_{\rm in,up,in}]
\nonumber \\ & ~
-\frac 12 i |T_{\frac12kh}|^2 |R_{\frac12kh}|^2 f_{\rm up}(h) \Lambda_{{\rm up,up},k}(h) 
& [2_{\rm up,in,up}]
\nonumber \\ & ~
+\frac 12i |T_{\frac12kh}|^2 T_{\frac12kh} R_{\frac12kh}^\ast f_{\rm up}(h) \Lambda_{{\rm up,in},k}(h)
& [2_{\rm up,up,in}]
\nonumber \\ & ~
-\frac12 i R_{\frac12,kh}^\ast T_{\frac12,kh} f_{\rm up}(h) \Lambda^\ast_{{\rm in,up},k}(h)
& [3]
\nonumber \\ & ~
-\frac12 i R_{\frac12,kh}^\ast T_{\frac12,kh} f_{\rm up}(h) {\cal E}^\ast_{{\rm in,up},k}(h)
& [4]
\nonumber \\ & ~
+\frac12 i R_{\frac12,kh}^\ast T_{\frac12,kh} f_{\rm up}(h) \Lambda^\ast_{{\rm in,up},k}(h)
\Bigr\}.& [5]
\label{eq:B1-expand}
\end{align}
The numbers on the right-hand side indicate which terms in Eq.~(\ref{eq:B1}), along with the ``in'' and ``up'' subscripts in the summation, that are being broken down. The $6^{\text{th}}$ term in Eq.~(\ref{eq:B1}) does not contribute to the imaginary part, and the $7^{\text{th}}$ and $8^{\text{th}}$ term do not have singularities of the $\eta^{-1}$ form. Finally, simplifying our expression by substituting Eq.~(\ref{eq:E-SUB}) for ${\cal E}_{XX'k}(h)$
provides a mass cancellation resulting in 
\begin{equation}
{\mathbb V}B_1(k,h)|_{\eta^{-1}} = 0.
\end{equation}

Having shown that there are no singular terms in $B_0$ and $B_1$ that need to be accounted for, we are justified in modifying the integrals to keep only the non-singular parts ${\mathbb U}B_0$ and ${\mathbb U}B_1$. These integrals are
\begin{align}
B_{0}(k,h) =& \Im \int_0^\infty \frac{dh'}{2\pi} \Biggl\{ \sum_{XX'Y}
w_{Xkh}w^\ast_{Ykh} \frac{{\cal A}_{XX'k}(h,h'){\cal A}^\ast_{YX'k}(h,h')}{(h-h'+i\eta)(h-h'+i\epsilon)}
\left[f_Y(h) f_{X'}(h') - \frac{f_Y(h) + f_{X'}(h')}{2} \right]
\nonumber \\ &~~
+ \sum_{XX'Y} w_{Xkh} w_{Ykh}^\ast {\cal I}^{(2)}_{XX'k}(h,h') {\cal I}^{(2)\ast}_{YX'k}(h,h')
\left[ \frac{1-f_Y(h)-f_{X'}(h')}{2}
+ f_Y(h) f_{X'}(h')
\right]
\Biggr\}
\label{eq:B0u}
\end{align}
and
\begin{align}
B_{1}(k,h) =& \Im \int_0^\infty \frac{dh'}{2\pi} \Biggl\{ \sum_{XX'Y} w_{Xkh} w_{Ykh}^\ast \frac{{\cal A}_{XX'k}(h,h'){\cal A}^\ast_{YX'k}(h,h')}{(h-h'+i\eta)(h-h'+i\epsilon)^2} \left[f_Y(h) - f_{X'}(h') \right]
\nonumber \\
& + \sum_{XX'Y} w_{Xkh} w_{Ykh}^\ast {\cal I}^{(2)}_{XX'k}(h,h'){\cal I}^{(2)\ast}_{YX'k}(h,h') \frac{f_Y(h)+f_{X'}(h')-1}{h+h'}
\Biggr\}.
\label{eq:B1u}
\end{align}

\subsection{Simplification of the $B_0$ term}
\label{sec:B0_derivation}

This section expresses Eq.~(\ref{eq:B0u}) in a form suitable for numerical computation: to the extent possible, we write the singular integrals in terms of derivatives of the transmission probability ${\mathbb T}_{\frac12 kh}$ with respect to semi-classical quantities, such as the ``atomic number'' $Z$ of the black hole, and derivatives of the electron phase space density. Due to the structure of the integrand, which contains two poles in the vicinity of $h'\approx h$, we expect $B_0(k,h)$ to contain only first derivatives of ${\mathbb T}_{\frac12 kh}$ and $f_{\rm up}(h)$.

On inspection of Eq.~(\ref{eq:B0u}), the numerator contributions of $w_{Xkh}w_{Ykh}^\ast {\cal A}_{XX'k}(h,h'){\cal A}^\ast_{YX'k}(h,h')$ are explicitly real when $X=Y$, but contain an imaginary component otherwise. Physically, the only nonzero terms in $B_0$ and the outgoing emission spectrum, due to the self-interaction of the electron field,  correspond to the interference of reflected (in the ``in'' basis) and transmitted electron modes (in the ``up'' basis) across the black hole's potential barrier. This enables us to use Eq.~(\ref{eq:PPII}), which breaks the singular integral into a principal part of the integral and a (half-)residue term capturing the behavior near $h'=h$:
\begin{eqnarray}
    B_{0}(k,h) &=& {\mathbb P}\int_0^\infty \frac{dh'}{2\pi} \Im \sum_{XX'Y}
w_{Xkh}w^\ast_{Ykh} \frac{{\cal A}_{XX'k}(h,h'){\cal A}^\ast_{YX'k}(h,h')}{(h-h')^2}
\left[f_Y(h) f_{X'}(h') - \frac{f_Y(h) + f_{X'}(h')}{2} \right] \nonumber\\
&& + \Im \int_0^\infty \frac{dh'}{2\pi} \sum_{XX'Y} w_{Xkh} w_{Ykh}^\ast {\cal I}^{(2)}_{XX'k}(h,h') {\cal I}^{(2)\ast}_{YX'k}(h,h')
\left[ \frac{1-f_Y(h)-f_{X'}(h')}{2}
+ f_Y(h) f_{X'}(h') \right]
\nonumber \\
&& + \frac{1}{2} \Re \frac{\partial}{\partial h'} \sum_{XX'Y}
w_{Xkh}w^\ast_{Ykh} {\cal A}_{XX'k}(h,h'){\cal A}^\ast_{YX'k}(h,h')
\left[f_Y(h) f_{X'}(h') - \frac{f_Y(h) + f_{X'}(h')}{2} \right] \Bigg\vert_{h'=h}.
\label{eq:B0-int-temp1}
\end{eqnarray}
The integrands are only imaginary when $X \neq Y$, and by grouping terms carefully, the two integrals in Eq.~(\ref{eq:B0-int-temp1}) can be simplified. The derivative (residue) term can be expanded using the product rule and then Eq.~(\ref{eq:E-SUB}) to express the derivatives of the ${\cal A}$'s in terms of the $\Lambda$'s. After a few lines of algebra, the result is
\begin{eqnarray}
    B_{0}(k,h) &=& - {\mathbb P}\int_0^\infty \frac{dh'}{2\pi} \frac{\Im \sum_{X'} R_{\frac12 kh}T^\ast_{\frac12 kh} {\cal A}_{\rm{in},X',k}(h,h'){\cal A}^\ast_{\rm{up},X',k}(h,h')  }{(h-h')^2} \left[\frac12- f_{X'}(h') \right]f_{\rm up}(h) \nonumber\\
&& - \int_0^\infty \frac{dh'}{2\pi} \Im \sum_{X'} R_{\frac12 kh}T^\ast_{\frac12 kh} {\cal I}^{(2)}_{{in},X'k}(h,h') {\cal I}^{(2)\ast}_{{\rm up},X'k}(h,h')
\left[\frac12- f_{X'}(h') \right]f_{\rm up}(h)
\nonumber \\
&& 
+ \left[\frac14
|T_{\frac12 kh}|^2 \frac{d|T_{\frac12 kh}|^2}{dh} + |R_{\frac12 kh}|^2 \Im\left\{ R_{\frac12 kh}T^\ast_{\frac12 kh} \Lambda_{\rm{in,up},k}(h,h) \right\} \right] \frac{f_{\rm up}(h)}{2} 
\nonumber\\
&&  + \left[-\frac12 |T_{\frac12 kh}|^2 \frac{d|T_{\frac12 kh}|^2}{dh} + |T_{\frac12 kh}|^2 \Im\left\{ R_{\frac12 kh}T^\ast_{\frac12 kh} \Lambda_{\rm{in,up},k}(h,h) \right\} \right] \frac{[1-f_{\rm up}(h)]f_{\rm up}(h)}{2} 
\nonumber\\
&& - \frac14 |T_{\frac12 kh}|^2\frac{df_{\rm up}}{dh} \left[ 1 - 2|T_{\frac12 kh}|^2f_{\rm up}(h)  \right].
\label{eq:B0-int-temp2}
\end{eqnarray}
The $1^{\text{st}}$ line corresponds to the principal part of the integral and describes the interference of reflected and transmitted electron modes of different energies; the $2^{\text{nd}}$ line details the effect that pair-production and pair-annihilation would produce; the $3^{\text{rd}}$ and $4^{\text{th}}$ lines correspond to the interference of reflected and transmitted modes \textit{of the same energy} and may be written in terms of derivatives of the transmission probability; finally, the $5^{\text{th}}$ line details how the self-interaction of the electron field alters the electron phase space density.

To make a direct comparison between our results and the semi-classical emission spectrum derived in Appendix~\ref{sec:sc_terms}, we now express $B_0$ in terms of semi-classical derivatives of the transmission probability. Terms without a copy of $f_{X'}(h')$ simplify using Eq.~(\ref{eq:Z-der1}) for $\partial_Z{\mathbb T}_{\frac12kh}(0)$ and Eq.~(\ref{eq:dyt-final}) for $\partial_Y{\mathbb T}_{\frac12kh}(0)$ leading to
\begin{eqnarray}
-\frac{\alpha}{M}  B_0(k,h)& =&
\partial_Z{\mathbb T}_{\frac12kh}(0) f_{\rm up}(h)
-\frac12 \partial_Y{\mathbb T}_{\frac12kh}(0) f_{\rm up}(h)
+ \frac{\alpha}{4M} {\mathbb T}_{\frac12kh} \frac{df_{\rm up}(h)}{dh} 
\nonumber \\ &&
-\frac12{\mathbb T}_{\frac12kh} [\partial_Z{\mathbb T}_{\frac12kh}(0)] [f_{\rm up}(h)]^2 - \frac{\alpha}{2M} {\mathbb T}_{\frac12kh}^2 f_{\rm up}(h) \frac{df_{\rm up}(h)}{dh} - \frac{\alpha}{4M} {\mathbb T}_{\frac12 kh} \frac{d{\mathbb T}_{\frac12 kh}}{dh} [f_{\rm up}(h)]^2 
\nonumber \\ &&
-\frac{\alpha}M f_{\rm up}(h) 
{\mathbb P}\int_0^\infty \frac{dh'}{2\pi} 
\Im \Bigl\{ R_{\frac12kh}T_{\frac12kh}^\ast 
\Bigl[ {\cal I}^{(1)}_{{\rm in,up},k}(h,h') {\cal I}^{(1)\ast}_{{\rm up,up},k}(h,h')
\nonumber \\ 
&&~~~~~~~~~~~~~~~~~~~~~
+ {\cal I}^{(2)}_{{\rm in,up},k}(h,h') {\cal I}^{(2)\ast}_{{\rm up,up},k}(h,h') \Bigr] \Bigr\}
f_{\rm up}(h'),
\label{eq:amb0}
\end{eqnarray}
which may be further simplified by substituting Eq.~(\ref{eq:dyt-final2}) for ${\mathbb T}_{\frac12 kh} d{\mathbb T}_{\frac12 kh}/dh$. Therefore, $B_0$ can be expressed purely in terms of $\partial_Z{\mathbb T}_{\frac12 kh}$, $\partial_Y {\mathbb T}_{\frac12 kh}$, and $df_{\rm up}/dh$ except for the principal part of the integral which is still expressed in our QFT formalism. This is given by

\begin{eqnarray}
-\frac{\alpha}{M} B_0(k,h) &=&
\partial_Z{\mathbb T}_{\frac12kh}(0) f_{\rm up}(h)
-\frac12 \partial_Y{\mathbb T}_{\frac12kh}(0) f_{\rm up}(h)
+ \frac{\alpha}{4M} {\mathbb T}_{\frac12kh} \frac{df_{\rm up}(h)}{dh}
-\frac12{\mathbb T}_{\frac12kh} [\partial_Z{\mathbb T}_{\frac12kh}(0)] [f_{\rm up}(h)]^2 
 \nonumber \\ &&
- \frac{\alpha}{2M} {\mathbb T}_{\frac12kh}^2 f_{\rm up}(h) \frac{df_{\rm up}(h)}{dh} + [\partial_Y {\mathbb T}_{\frac12 kh}(0)] [f_{\rm up}(h)]^2 - [\partial_Z {\mathbb T}_{\frac12 kh}(0)] [f_{\rm up}(h)]^2
\nonumber  \\ &&
-\frac{\alpha}M 
{\mathbb P}\int_0^\infty \frac{dh'}{2\pi} 
\Im \Bigl\{ R_{\frac12kh}T_{\frac12kh}^\ast
\Bigl[ {\cal I}^{(1)}_{{\rm in,up},k}(h,h') {\cal I}^{(1)\ast}_{{\rm up,up},k}(h,h') 
\nonumber \\ &&~~~~~~~~~~~~
+ {\cal I}^{(2)}_{{\rm in,up},k}(h,h') {\cal I}^{(2)\ast}_{{\rm up,up},k}(h,h') \Bigr] \Bigr\}
f_{\rm up}(h) \left[ f_{\rm up}(h') - f_{\rm up}(h)  \right]
\nonumber \\
&&
+\frac{\alpha}M 
{\mathbb P}\int_0^\infty \frac{dh'}{2\pi} 
\Im \Bigl\{ R_{\frac12kh}T_{\frac12kh}^\ast
\Bigl[ {\cal I}^{(1)}_{{\rm in,in},k}(h,h') {\cal I}^{(1)\ast}_{{\rm up,in},k}(h,h')
+ {\cal I}^{(2)}_{{\rm in,in},k}(h,h') {\cal I}^{(2)\ast}_{{\rm up,in},k}(h,h') \Bigr] \Bigr\}[f_{\rm up}(h)]^2.~~~~~
\label{eq:B_0_final}
\end{eqnarray}
Comparing Eq.~(\ref{eq:B_0_final}) with our replication of the semi-classical emission spectrum, given by Eq.~(\ref{eq:sc_spectra}), there are terms in $B_0(k,h)$ that replicate the semi-classical emission spectrum (terms without $\langle Z^2 \rangle$). Expressing $B_0$ as
\begin{equation}
    B_0(k,h) = B_0(k,h) \vert_{\rm sc} + B_0(k,h) \vert_{\rm QFT}
\end{equation}
we wish to isolate which terms are semi-classical in origin and which terms can only be understood within the context of QFT. The semi-classical contribution is given by
\begin{equation}
    - \frac{\alpha}{M} B_0(k,h) \vert_{\rm sc} = \left[\partial_Z {\mathbb T}_{\frac12 kh}(0) \right] f_{\rm up}(h) + \frac{\alpha}{2M} {\mathbb T}_{\frac12 kh}(0) \frac{df_{\rm up}}{dh}
\label{eq:B0_sc}
\end{equation}
while our QFT correction is given by
\begin{equation}
\begin{split}
    - \frac{\alpha}{M} B_0(k,h) \vert_{\rm QFT} =& - \frac{1}{2} [\partial_Y {\mathbb T}_{\frac12 kh}(0)] f_{\rm up}(h) \left[1- 2f_{\rm up}(h) \right] -\frac{\alpha}{4M} {\mathbb T}_{\frac12 kh}(0) \left[1+2{\mathbb T}_{\frac12 kh}(0)f_{\rm up}(h)\right] \frac{df_{\rm up}}{dh} \\
    \ &- [\partial_Z {\mathbb T}_{\frac12 kh}(0)] \left[1 + \frac{1}{2} {\mathbb T}_{\frac12 kh}(0) \right] [f_{\rm up}(h)]^2 \\
    \ & -\frac{\alpha}M 
{\mathbb P}\int_0^\infty \frac{dh'}{2\pi} 
\Im \Bigl\{ R_{\frac12kh}T_{\frac12kh}^\ast
\Bigl[ {\cal I}^{(1)}_{{\rm in,up},k}(h,h') {\cal I}^{(1)\ast}_{{\rm up,up},k}(h,h') \\
\ & ~~~~~~~~~~~~~~~~~~~~~~~~~~~~~~~~+ {\cal I}^{(2)}_{{\rm in,up},k}(h,h') {\cal I}^{(2)\ast}_{{\rm up,up},k}(h,h') \Bigr] \Bigr\}
f_{\rm up}(h) \left[ f_{\rm up}(h') - f_{\rm up}(h)  \right] \\
\ & +\frac{\alpha}M 
{\mathbb P}\int_0^\infty \frac{dh'}{2\pi} 
\Im \Bigl\{ R_{\frac12kh}T_{\frac12kh}^\ast
\Bigl[ {\cal I}^{(1)}_{{\rm in,in},k}(h,h') {\cal I}^{(1)\ast}_{{\rm up,in},k}(h,h')
+ {\cal I}^{(2)}_{{\rm in,in},k}(h,h') {\cal I}^{(2)\ast}_{{\rm up,in},k}(h,h') \Bigr] \Bigr\}[f_{\rm up}(h)]^2.
 \end{split}
\label{eq:B0_QFTterms}
\end{equation}
In this separation, $B_0|_{\rm sc}$ (\ref{eq:B0_sc}) represents the attraction of the electron to the positive charge left behind (it is the change of spectrum when $Z$ is incremented by 1, including the $df_{\rm up}/dh$ term associated with the change in potential at the horizon), and is equivalent to the calculation of \citet{PhysRevD.16.2402}. The remainder of the terms are in $B_0|_{\rm QFT}$ (Eq.~\ref{eq:B0_QFTterms}). Most notable are the $\partial_Y{\mathbb T}$ term that describes how the self-energy of the electron perturbs the outgoing emission spectra, and the integral over $dh'$ that represents the exchange term mediated by the 0L component of the electrostatic interaction. But note that due to the handling of the singularity at $h'=h$, there are many other terms as well. Multiple forms of this equation are possible, but the principal part form is best suited to numerical computation.

\subsection{Separation of the parts of $B_1$}

In this section, we express Eq.~(\ref{eq:B1}) in terms of derivatives of the transmission probability ${\mathbb T}_{\frac12 kh}$, with respect to the semi-classical charge of the black hole, and derivatives of the electron phase space density  $f_{\rm up}(h)$. Unlike $B_0(k,h)$ which contains terms that can only be explained within the context of our QFT formalism, given by Eq.~(\ref{eq:B0_QFTterms}), $B_1(k,h)$ replicates the semi-classical prediction completely. There are no additional terms that we need to consider.

We begin our derivation by noting that Eq.~(\ref{eq:B1u}) has two types of contributions: terms that contain $f_Y(h)$ and terms that do not (with $f_{X'}(h')$ and 1). The terms without $f_Y(h)$ can be written as
\begin{equation}
B_1(k,h)\vert_{{\rm not}~f_Y(h)} = -\Im \int_0^\infty \frac{dh'}{2\pi} \frac{\left| \sum_X w_{Xkh} {\cal A}_{X,{\rm up},k}(h,h') \right|^2}{(h-h'+i\upsilon)^3} f_{\rm up}(h')
\label{eq:B1_notfY}
\end{equation}
(the ``other'' terms containing ${\cal I}^{(2)}$ are manifestly real and do not contribute). Since the integral is non-singular as long as the poles are located on the same side of the complex plane, we remove the distinction between the $\epsilon$ and $\eta$ as their distance from the real axis and the order of limits does not matter.    Using Eq.~(\ref{eq:Zh3}), we find that
\begin{equation}
B_1(k,h)\vert_{{\rm not}~f_Y(h)} = \frac14 \partial_{h'}^2 \Bigl[ \Bigl| \sum_X w_{Xkh} {\cal A}_{X,{\rm up},k}(h,h') \Bigr|^2 f_{\rm up}(h') \Bigr]_{h'=h},
\end{equation}
which can be expanded further by using the product rule and noting that from the special values of ${\cal A}$ (Eq.~\ref{eq:A-val}):
\begin{equation}
\sum_X w_{Xkh} {\cal A}_{X,{\rm up},k}(h,h)
= iT_{\frac12,kh}
~~~\Rightarrow~~~
\Bigl|\sum_X w_{Xkh} {\cal A}_{X,{\rm up},k}(h,h)\Bigr|^2
= {\mathbb T}_{\frac12kh}(0).
\label{eq:coef-0}
\end{equation}

Including terms in Eq.~(\ref{eq:B1u}) that contain $f_Y(h)$, and using Eq.~(\ref{eq:aeq}) to express the derivatives of ${\cal A}_{XX'k}(h,h')$ evaluated at $h=h'$,
we find
\begin{eqnarray}
B_1(k,h) &=& \frac14 |T_{\frac12kh}|^2 \frac{d^2f_{\rm up}}{dh^2}
+ \Im \Bigl[ T^\ast_{\frac12kh} \sum_X w_{Xkh}{\cal E}_{X,{\rm up},k}(h) \Bigr]
\frac{df_{\rm up}}{dh}
\nonumber \\ &&
+ \Biggl\{
\frac12 \Im \Bigl[ T^\ast_{\frac12kh} \sum_X w_{Xkh} Q_{X,{\rm up},k}(h)
\Bigr]
 +
\frac12 \Bigl|  \sum_X w_{Xkh} {\cal E}_{X,{\rm up},k}(h)  \Bigr|^2
\nonumber \\ &&
~~ + \Im \int_0^\infty \frac{dh'}{2\pi} \sum_{XX'} T^\ast_{\frac12kh} w_{Xkh} \Bigl[\frac{{\cal A}_{XX'k}(h,h') {\cal A}^\ast_{{\rm up},X'k}(h,h')}{(h-h'+i\upsilon)^3}
+ \frac{{\cal I}^{(2)}_{XX'k}(h,h') {\cal I}^{(2)\ast}_{{\rm up},X'k}(h,h')}{h+h'} \Bigr]
\Biggr\}f_{\rm up}(h).
\label{eq:B1-expanded-A}
\end{eqnarray}
where the second term in Eq.~(\ref{eq:B1-expanded-A}) can be further simplified using
\begin{equation}
    \Im \Bigl[T^\ast_{\frac12 kh}\sum_X w_{Xkh} {\cal E}_{X,{\rm up},k}(h) \Bigr] = - \Im \left[ |T_{\frac12 kh}|^2 \Lambda_{{\rm up,up},k}(h,h) + R_{\frac12 kh}T^\ast_{\frac12 kh} \Lambda_{{\rm in,up},k}(h,h) \right] = \frac{M}{\alpha} \partial_Z {\mathbb T}_{\frac12 kh}(0).
\end{equation}
Here we expressed the $\cal E$-terms using Eq.~(\ref{eq:E2}) which simplifies to a linear combination of $\Lambda$-terms. Since $\Lambda_{{\rm up,up},k}(h,h)$ is real, the only nonzero contribution comes from the $\Lambda_{{\rm in, up},k}(h,h)$ term which we relate to $\partial_Z {\mathbb T}_{\frac12 kh}(0)$ using Eq.~(\ref{eq:Z-der1}). Likewise, the $2^{\text{nd}}$ and $3^{\text{rd}}$ lines of Eq.~(\ref{eq:B1-expanded-A}) are related to $\partial_Z^2 {\mathbb T}_{\frac12 kh}(0)$ by Eq.~(\ref{eq:Z2-der}) (the derivation is rather technical, so we have placed it in Appendix~\ref{app:T2}). Finally, the stochastic charge correction given by our QFT approach is
\begin{equation}
\frac{\alpha^2}{2M^2}
B_1(k,h) = \frac{\alpha^2}{8 M^2} {\mathbb T}_{\frac12kh}(0) \frac{d^2f_{\rm up}}{dh^2}
+ \frac{\alpha}{2M} [\partial_Z{\mathbb T}_{\frac12kh}(0)] 
\frac{df_{\rm up}}{dh}
+ \frac{1}{2}[\partial_Z^2{\mathbb T}_{\frac12kh}(0)] f_{\rm up}(h),
\label{eq:B1-simplify}
\end{equation}
which, despite being  an $\mathcal{O}(\alpha^2)$ effect, will produce an $\mathcal{O}(\alpha)$ correction to the outgoing electron emission spectrum due to $\langle \hat{\cal Z}^2 \rangle \sim \mathcal{O}(\alpha^{-1})$.
Comparing Eq.~(\ref{eq:B1-simplify}) with Eq.~(\ref{eq:sc_spectra}), the stochastic charge correction derived by our QFT approach is identical to the result derived by \citet{PhysRevD.16.2402} using semi-classical arguments. This establishes Conjecture \#2.

\section{Numerical Calculations}
\label{sec:numerics}

\subsection{Parameters}
\label{subsec:params}


We computed the numerical results of the stochastic charge effect correction for PBH masses ($M$) of $1\times10^{21}$, $2\times10^{21}$, $4\times10^{21}$ and $8\times10^{21}$ Planck masses. For each mass, energies are expressed in units of the corresponding Hawking temperature, $T_{\rm H}=1/(8\pi M)$. The outgoing electron energy ($h$) grid is 200 linearly spaced points with a spacing of $0.1T_H$ across the range $0.1T_H$ to $20T_H$ for each black hole mass. The $k$ grid has extremal values of $-5$ and 5 and a spacing of 1 (excluding $k=0$). The convergence of the angular-momentum truncation at $|k|=5$ is discussed in Appendix \ref{sec:rk_error}. Parameters specific to the evaluation of the radial overlap integrals and transmission probabilities are introduced in the following subsections respectively. 

\subsection{Terms involving ${\cal I}$-integrals}
\label{subsec:Bowen's part}

The perturbation to the primary electron emission spectrum Eq.~(\ref{eq:NHT}) can be separated into a stochastic component $B_1(k,h)$ and a self-interaction component $B_0(k,h)$. While most contributions to the corrections can be expressed through derivatives of the transmission probability, the final two terms of  Eq.~(\ref{eq:B_0_final}) for $B_0(k,h)$, are the only terms that require the explicit evaluation of the radial integrals $\mathcal{I}_{XX'k}^{(1)}(h,h')$ and $\mathcal{I}^{(2)}_{XX'k}(h,h')$, given by Eq.~(\ref{eq:I12-expression}).

A challenge in evaluating the ${\cal I}$-integrals is that the integrand in Eq.~(\ref{eq:I12-expression}) is does not converge to zero as $r_\star\rightarrow -\infty$, since $F$ and $G$ are oscillatory but $2M/r\rightarrow 1$. For ${\cal I}^{(1)}$, the product ··of these oscillations has a spatial frequency $h-h'$, resulting in a $1/(h-h')$ divergence in the integral over the negative infinite domain. This is handled analytically by the principal part, but the numerical evaluation of the integral still requires careful treatment. We partition the $\mathcal{I}$ integrals into a finite numerical region and an analytic tail region. This methodology allows us to fully utilize the electron radial wave function data meanwhile maintaining the control over the singularity induced by horizon. 

For the finite domain, we perform numerical integration using the radial grid parameters established in \citet{Koivu_2025}, in order to maintain consistency with our group's previous numerical evaluations and leverage the existing 
datasets for electron radial functions. The tortoise coordinate $r_\star$ is sampled over $240,000$ equally spaced points within the range $-70 M\le r_\star\le 2000 M$. For energy grid, we employ a double grid strategy for the energy dimensions $h$ and $h^\prime$ to balance resolution with computational efficiency. For $h^\prime$, we keep on a fine grid of 2000 equally spaced points ranging from $0.01T_H$ to $20T_H$, matching the resolution of the electron radial functions while for $h$ we use a grid of 200 points as discussed in Sec.~\ref{subsec:params}. High resolution is required here because the spectrum correction involves an integration over $h^\prime$, and we have verified that the 2000-point resolution is sufficiently fine for numerical convergence of the integrated correction terms. 

The treatment of the region $r_{\star} < -70 M$ requires a separate analytic approach because of the $1/(h-h^\prime)$ divergence of the $\mathcal{I}^{(1)}$ integrals as $h\rightarrow h^\prime$. To address this divergence, we evaluate an analytic tail of the integral by substituting the asymptotic forms of the electron radial wave functions Eq.~(\ref{eq:FG}) into the integrand, following section IV of \citet{PhysRevD.107.045004}, and introducing a regularization parameter $\epsilon = 10^{-5} T_H$.
The explicit forms of the radial electron functions near the horizon are:
\begin{equation}\label{eq:FG}
\begin{pmatrix}F_\text{up}\\G_\text{up}\end{pmatrix}\xrightarrow[r_*\to-\infty]{}
\begin{cases}
\begin{pmatrix}\sqrt{h}\\ -i\sqrt{h}\end{pmatrix}e^{ihr_\star}
-R^{*}_{\frac{1}{2},k,h}e^{2i\arg(T_{1/2,k,h})}\begin{pmatrix}\sqrt{h}\\ i\sqrt{h}\end{pmatrix}e^{-ihr_\star},
&  (h>\mu)\ \\[4pt]
\begin{pmatrix}\sqrt{h}\\ -i\sqrt{h}\end{pmatrix}e^{ihr_\star}
+e^{2i\delta_{1/2,k,h}}\begin{pmatrix}\sqrt{h}\\ i\sqrt{h}\end{pmatrix}e^{-ihr_\star},
&  (0<h<\mu)\ .
\end{cases}
\end{equation}
and
\begin{equation}\label{eq:near-horizon-asymptotics}
\begin{aligned}
\binom{F_\text{in}}{G_\text{in}}
&\xrightarrow[r_*\to-\infty]{}
T_{\frac12,k,h}\binom{\sqrt{h}}{i\sqrt{h}}\,e^{-ih r_*},
\qquad (h>\mu)\ .
\end{aligned}
\end{equation}
(There are no ``in'' modes at $h<\mu$.)
Since $\mathcal{I}^{(2)}$ integrals do not exhibit the singular behavior, they are integrated purely through numerical summation over $-70M \le r_{\star} \le 2000 M$.

The final stage involves the integration over $h'$ appearing in the final two terms of Eq.~(\ref{eq:B_0_final}), which must be evaluated as a Cauchy principal part due to the $1/(h-h')$ divergence behavior of $\mathcal{I}^{(1)}$ integrals. On the discrete $h$ and $h'$ grid we implement the principal part prescription by excluding the diagonal point where $h=h^\prime$ and evaluating as the symmetric sum about the pole. For each $h$, we pair the contributions at $h^{\prime}=h \pm j \Delta h'$ to cancel the singular part. The evaluation error is dominated by the discrete-grid representation of the near-pole contribution. To estimate this bias, write the integrand near the pole as $\Phi(h,h')/(h-h')$, where $\Phi(h,h')$ is regular at $h=h'$. Let $x \equiv h'-h$ and expand near the pole:
\begin{equation}
\Phi(h, h+x)=\Phi_0(h)+\mathcal{O}(x), \quad \Phi_0(h) \equiv \Phi(h, h),
\end{equation}
so the leading near-pole term is $\Phi_0(h)/x$, which controls the principal-value bias. On our symmetric grid, the difference between the discrete sum and the corresponding continuum integral for each side of the pole are

\begin{eqnarray}
\Delta h^{\prime} \sum_{j=1}^J \frac{\Phi_0(h)}{j \Delta h^{\prime}}-\int_{\Delta h^{\prime}}^{(J+1) \Delta h^{\prime}} \frac{\Phi_0(h)}{x} d x &=& \Phi_0(h)\,[H_J-\ln (J+1)]\, \underset{J \rightarrow \infty}{\longrightarrow} \gamma_E \Phi_0(h) ~~~{\rm and} \nonumber \\
\Delta h^{\prime} \sum_{j=-J}^{-1} \frac{\Phi_0(h)}{j \Delta h^{\prime}}-\int_{-(J+1) \Delta h^{\prime}}^{-\Delta h^{\prime}} \frac{\Phi_0(h)}{x} d x &=& -\Phi_0(h)\,[H_J-\ln (J+1)] \underset{J \rightarrow \infty}{\longrightarrow}-\gamma_E \Phi_0(h),
\end{eqnarray}
where $H_J=\sum_{j=1}^J 1 / j$ and $\gamma_E \simeq 0.577 $ is the Euler–Mascheroni constant. Since residuals on two sides of the pole carry opposite signs and cancel exactly upon summation, this cancellation ensures that the primary discretization bias vanishes. 

\subsection{Terms involving transmission probabilities} \label{subsec:Cara's part}

All of the terms not involving ${\cal I}$ integrals are combinations of the transmission probability ${\mathbb T}_{\frac12 kh}$ and its first and second derivatives, and the unperturbed electron phase space density $f_{\rm up}(h)$ and its first and second derivatives. The derivatives with respect to $h$ of the phase space density can be calculated analytically (\S \ref{subsubsec:fup}). The transmission probabilities are calculated numerically (\S \ref{subsubsec:integrator}). Derivatives of the transmission probability with respect to $Z$ and $Y$ are calculated numerically (\S \ref{subsubsec:derivatives}).

\subsubsection{Phase space density derivatives}
\label{subsubsec:fup}

With $f_{up}(h)$ defined as in Table \ref{tab:math_terms_from_first_paper}, we can determine its first and second derivatives analytically:
\begin{equation}
    f_{up}(h)=\frac{1}{e^{8\pi M h}+1}
    ~~\rightarrow~~
    f_{up}'(h)=-\frac{2\pi M}{\cosh^2(4\pi M h)}
    ~~\rightarrow~~
    f_{up}''(h)=\frac{16 \pi^2 M^2 \tanh(4 \pi M h)}{\cosh^2(4 \pi M h)};
\end{equation}
these are then implemented numerically with the \texttt{NumPy} \texttt{exp}, \texttt{tanh} and \texttt{cosh} functions \cite{2020Natur.585..357H}.

\subsubsection{Transmission probability integrator}\label{subsubsec:integrator}


The next ingredient we need is to numerically compute the first quantized transmission probability ${\mathbb T}_{\frac12kh}$ through the angular momentum barrier as a function of the charge $Z$ on the black hole and the amplitude $Y$ of the $1/r^2$ term in the potential. This is calculated through the radial equation (equation 39) from \citet{1957RvMP...29..465B}, with appropriate extra terms proportional to $Z$ and $Y$:
\begin{equation}
 \left(h- \frac{\alpha Z}{r} -\frac{2\alpha M Y}{r^2}\right)  {\boldsymbol\xi}
= 
\left( \begin{array}{cc} \mu\nu & \frac kr\nu + \partial_{r_\star} \\ \frac kr\nu -\partial_{r_\star} & -\mu\nu \end{array}\right)
{\boldsymbol\xi},
~~~{\rm where}~~~ {\boldsymbol\xi}\equiv \left(\begin{array}cF \\ G\end{array}\right)
~~~{\rm and}~~~
\nu \equiv \sqrt{1-\frac{2M}r}
.
\label{eq:energy_matrix}
\end{equation}


Then Eq.~(\ref{eq:energy_matrix}) can be rearranged and used to define the $2\times 2$ matrix ${\cal \Tilde{D}}_{r,Z,Y,k,h}$ as follows
\begin{equation} \label{eq:Dmat}
    \frac{\partial\boldsymbol \xi}{\partial r} = \frac{1}{\nu} \left( \begin{array}{cc} \frac{k}{r} & -\mu+\frac{1}{\nu}\left(h-\frac{\alpha Z}{r}-\frac{2 \alpha M Y}{r^2}\right) \\ -\mu-\frac{1}{\nu}\left(h-\frac{\alpha Z}{r}-\frac{2 \alpha M Y}{r^2}\right) &  -\frac{k}{r} \end{array}\right) {\boldsymbol\xi} \equiv {\cal \Tilde{D}}_{r,Z,Y,k,h}\, {\boldsymbol\xi}.
\end{equation}
The following leapfrog integrator can be used to numerically determine ${\boldsymbol\xi}$ values on consecutive $r_i$ grid points:
\begin{equation}
    {\boldsymbol\xi}_{i+1}=\left({\mathbb I}+\frac{\Delta r}{2} {\cal \Tilde{D}}_{i+\frac12}\right)\left({\mathbb I}-\frac{\Delta r}{2} {\cal \Tilde{D}}_{i+\frac12}\right)^{-1} {\boldsymbol\xi}_{i} \equiv {\cal \Tilde{T}}_{i+\frac12} {\boldsymbol\xi}_{i} ,
    \label{eq:tildeT definition}
\end{equation}
where ${\cal \Tilde{D}}_{i+\frac12}$ is ${\cal \Tilde{D}}_{r,Z,Y,k,h}$ calculated at the arithmetic mean of $r_i$ and $r_{i+1}$ for given $Z$, $Y$, $k$ and $h$ values, and ${\cal \Tilde{T}}_{i+\frac12}$ can be understood in a similar manner. This integrator is chosen because it maintains the determinant of ${\cal \Tilde{T}}$ at exactly 1 (since $\cal \Tilde{D}$ is traceless) --- it thus has the general advantages of symplectic integrators in Hamiltonian mechanics \cite{1990Nonli...3..231C}. 

In order to integrate inwards (a transmission coefficient could be done either way), we can define
\begin{equation}
    {\cal T}_{i}\equiv{\cal \Tilde{T}}_i^{-1}=\left({\mathbb I}+\frac{\Delta r}{2} {\cal D}_{i}\right)\left({\mathbb I}-\frac{\Delta r}{2} {\cal D}_{i}\right)^{-1}
    \label{eq:T definition}
\end{equation}
where ${\cal D}_{i}\equiv -{\cal \Tilde{D}}_{i}$ and then reorder Eq.~(\ref{eq:tildeT definition}) as ${\boldsymbol\xi}_{i-1}={\cal T}_{i-\frac12} {\boldsymbol\xi}_{i}$,
where ${\cal T}_{i-\frac12}$ is evaluated at the arithmetic mean of $r_{i-1}$ and $r_i$.
We can use this result iteratively and find that
\begin{equation}
    {\boldsymbol\xi}_{n}={\cal T}_{n+\frac12}{\cal T}_{n+\frac{3}{2}}...{\cal T}_{\text{max}-\frac{3}{2}} {\cal T}_{\text{max}-\frac12} {\boldsymbol\xi}_{\text{max}} \equiv {\cal T}_{\text{max} \rightarrow n} {\boldsymbol\xi}_{\text{max}},
\label{eq:xi}
\end{equation}
where ${\cal T}_{\text{max} \rightarrow n}$ is defined as the indicated matrix product.

We then use Eqs.~(\ref{eq:Dmat}), (\ref{eq:T definition}) and (\ref{eq:xi}) to iteratively compute ${\cal T}_{\text{max} \rightarrow n}$ on a logarithmically spaced grid ranging from $r_{\rm min} = (2+2.8\times 10^{-6})M$ to $r_{\rm max}=10^6M$ with $n_{r\text{-steps}}=1.81\times 10^5$ steps.
The errors from the limited range in $r$ are discussed in Appendix~\ref{sec:rk_error}.

If we set $\boldsymbol{\xi}_{\text{n}}$ equal to the upgoing asymptotic solution close to the horizon (determined in \cite{PhysRevD.107.045004}), we use Eq.~(\ref{eq:xi}) to calculate $\boldsymbol{\xi}_{\text{max}}$ at the point on our grid furthest away from the horizon using the \texttt{NumPy} \texttt{linalg.solve} function. Since the goal is to calculate transmission probability, the phase of each $\boldsymbol{\xi}$ is unimportant. We can then decompose $\boldsymbol{\xi}_\text{max}$ into its downgoing and outgoing components, and use the outgoing power to determine the transmission probability.

\subsubsection{Transmission Probability Derivatives}\label{subsubsec:derivatives}

The numerical derivatives of the transmission probability ${\mathbb T}_{\frac12kh}(Z,Y)$ are calculated using the five point finite difference method. Since all derivatives are to be evaluated where $Z$, $Y=0$, we calculated the values of the transmission probabilities at $Y,Z\in \{-2,-1,0,1,2\}$.
The first derivative with respect to $Z$ is therefore evaluated as 
\begin{equation}
    \partial_Z{\mathbb T}_{\frac12kh}(0,0)=\frac{{\mathbb T}_{\frac12kh}(-2,0) -8 {\mathbb T}_{\frac12kh}(-1,0) +8 {\mathbb T}_{\frac12kh}(1,0) - {\mathbb T}_{\frac12kh}(2,0)}{12}
\end{equation}
with error of order $(\alpha \Delta Z)^4$, and similarly for the first derivative with respect to $Y$.
The second derivative with respect to $Z$, also with error of order $(\alpha\Delta Z)^4$, is evaluated as 
\begin{equation}
    \partial^2_Z{\mathbb T}_{\frac12kh}(0,0)=\frac{-{\mathbb T}_{\frac12kh}(-2,0) +16 {\mathbb T}_{\frac12kh}(-1,0) -30 {\mathbb T}_{\frac12kh}(0,0)  +16 {\mathbb T}_{\frac12kh}(1,0) - {\mathbb T}_{\frac12kh}(2,0)}{12}.
\end{equation}


\subsection{Numerical Results}\label{subsec:results}


\subsubsection{Breakdown of Terms Related to $B_0$ and $B_1$}\label{subsubsec:B0}
The terms that constitute $B_0$ and $B_1$ are labeled in the order that they appear in Eq.~(\ref{eq:B_0_final}) and Eq.~(\ref{eq:B1-simplify}). Relevant coefficients and summations over $k$ are computed and included in all figures. The collection of terms contained within $B_0$ are labeled as

\begin{align*}
    t_{01} &= \sum_k \left(\frac{2j+1}{2\pi}\right) \partial_Z{\mathbb T}_{\frac12kh}(0) f_{\rm up}(h) ~~~~~~~~~~~~~~~~~~~~~~~ t_{02} = - \sum_k \left(\frac{2j+1}{2\pi}\right) \frac12 \partial_Y{\mathbb T}_{\frac12kh}(0) f_{\rm up}(h)\displaybreak[1]\\
    t_{03} &=  \sum_k \left(\frac{2j+1}{2\pi}\right) \frac{\alpha}{4M} {\mathbb T}_{\frac12kh} \frac{df_{\rm up}(h)}{dh} ~~~~~~~~~~~~~~~~~~~~~~~ t_{04} = - \sum_k \left(\frac{2j+1}{2\pi}\right) \frac12{\mathbb T}_{\frac12kh} [\partial_Z{\mathbb T}_{\frac12kh}(0)] [f_{\rm up}(h)]^2\displaybreak[1]\\
    t_{05} &= - \sum_k \left(\frac{2j+1}{2\pi}\right) \frac{\alpha}{2M} {\mathbb T}_{\frac12kh}^2 f_{\rm up}(h) \frac{df_{\rm up}(h)}{dh} ~~~~~~~~~~~~ t_{06} = \sum_k \left(\frac{2j+1}{2\pi}\right) [\partial_Y {\mathbb T}_{\frac12 kh}(0)] [f_{\rm up}(h)]^2\displaybreak[1]\\
    t_{07} &= - \sum_k \left(\frac{2j+1}{2\pi}\right) [\partial_Z {\mathbb T}_{\frac12 kh}(0)] [f_{\rm up}(h)]^2 \displaybreak[1]\\
    t_{08} &= - \sum_k \left(\frac{2j+1}{2\pi}\right) \frac{\alpha}M 
    {\mathbb P}\int_0^\infty \frac{dh'}{2\pi} 
    \Im \Bigl\{ R_{\frac12kh}T_{\frac12kh}^\ast
    \Bigl[ {\cal I}^{(1)}_{{\rm in,up},k}(h,h') {\cal I}^{(1)\ast}_{{\rm up,up},k}(h,h')
    + {\cal I}^{(2)}_{{\rm in,up},k}(h,h') {\cal I}^{(2)\ast}_{{\rm up,up},k}(h,h') \Bigr] \Bigr\}\\
    &\times f_{\rm up}(h) \left[ f_{\rm up}(h') - f_{\rm up}(h)  \right]\displaybreak[1]\\
    t_{09} &= \sum_k \left(\frac{2j+1}{2\pi}\right) \frac{\alpha}
    {\mathbb P}\int_0^\infty \frac{dh'}{2\pi} 
    \Im \Bigl\{ R_{\frac12kh}T_{\frac12kh}^\ast
    \Bigl[ {\cal I}^{(1)}_{{\rm in,in},k}(h,h') {\cal I}^{(1)\ast}_{{\rm up,in},k}(h,h') 
    + {\cal I}^{(2)}_{{\rm in,in},k}(h,h') {\cal I}^{(2)\ast}_{{\rm up,in},k}(h,h') \Bigr] \Bigr\}[f_{\rm up}(h)]^2 
\end{align*}

\begin{figure}
    \centering
    \includegraphics[trim={0.35in 1.4in 0.35in 1.5in}, clip, width=\linewidth]{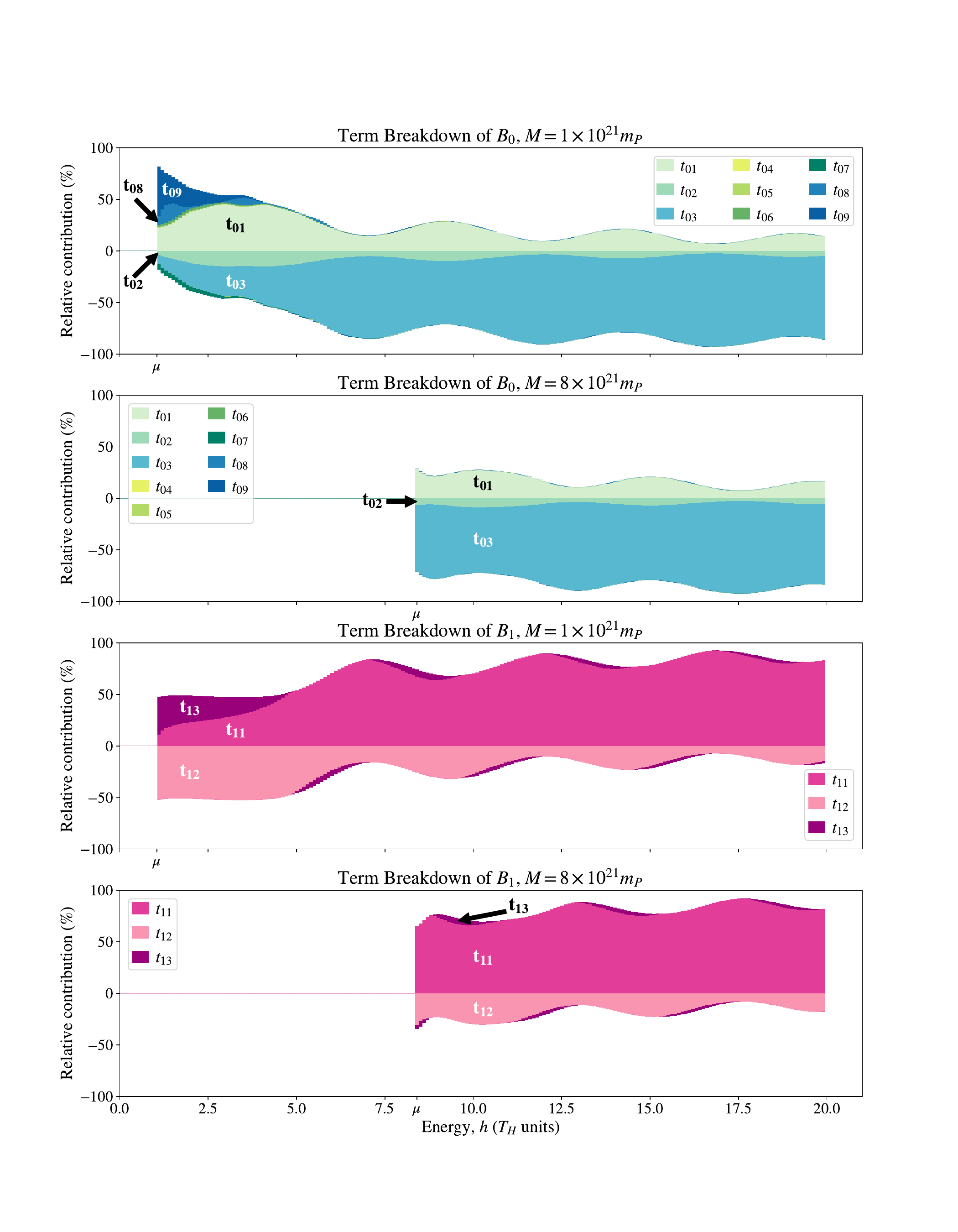}
    \caption{This figure shows the contribution of each term to the $\mathcal{O}(\alpha)$ correction to the electron number spectrum of two black hole masses, relative to the total contribution from $B_0$ terms (top 2 panels) and $B_1$ terms (bottom 2 panels). For each energy grid point, a bar composed of different colors is shown with a total bar length of 100\%. The length of each colored section of the bar represents the relative contribution of the term associated with that color at that energy, where positive contributions are shown above the 0\% relative contribution axis and negative contributions are shown below the 0\% relative contribution axis. Additional term labels are indicated directly on each panel for terms with significant contributions.}
    \label{fig:B0B1breakdownM1M8}
\end{figure}

while $B_1$ terms are labeled as
\begin{align*}
    t_{11} &= \sum_k (2j+1) \langle \hat{\cal Z}^2 \rangle \frac{\alpha^2}{16 \pi M^2} {\mathbb T}_{\frac12kh}(0) \frac{d^2f_{\rm up}}{dh^2} \displaybreak[1]\\
    t_{12} &= \sum_k (2j+1) \langle \hat{\cal Z}^2 \rangle \frac{\alpha}{4 \pi M} [\partial_Z{\mathbb T}_{\frac12kh}(0)] 
    \frac{df_{\rm up}}{dh}\displaybreak[1]\\
    t_{13} &= \sum_k (2j+1) \langle \hat{\cal Z}^2 \rangle \frac{1}{4 \pi}[\partial_Z^2{\mathbb T}_{\frac12kh}(0)] f_{\rm up}(h).
\end{align*}

In the low energy limit for lower masses, there are a large number of terms that contribute to the $B_0$ correction (see first panel of Fig. \ref{fig:B0B1breakdownM1M8}). The terms $t_{01}$ and $t_{09}$ contribute significantly in this limit, with additional moderate contributions from $t_{02}$, $t_{03}$, and $t_{08}$ and slight contributions from $t_{06}$ and $t_{07}$. The $B_0$ correction to the electron spectrum for higher masses (second panel of Fig. \ref{fig:B0B1breakdownM1M8}) and beyond the low energy limit for lower masses (first panel of Fig. \ref{fig:B0B1breakdownM1M8}) is dominated by contributions from $t_{03}$, with moderate contributions from $t_{01}$ and slight contributions from $t_{02}$.

Similarly, in the low energy limit for lower masses, the contribution to the $B_1$ correction is spread across its terms (see third panel of Fig. \ref{fig:B0B1breakdownM1M8}), with $t_{12}$ and $t_{13}$ contributing more significantly than $t_{11}$. For higher masses (fourth panel of Fig. \ref{fig:B0B1breakdownM1M8}) and higher energies (third panel of Fig. \ref{fig:B0B1breakdownM1M8}), $t_{11}$ becomes dominant with $t_{12}$ providing a moderate contribution and $t_{13}$ contributing only slightly. 

\begin{figure}
    \centering
    \includegraphics[width=\linewidth]{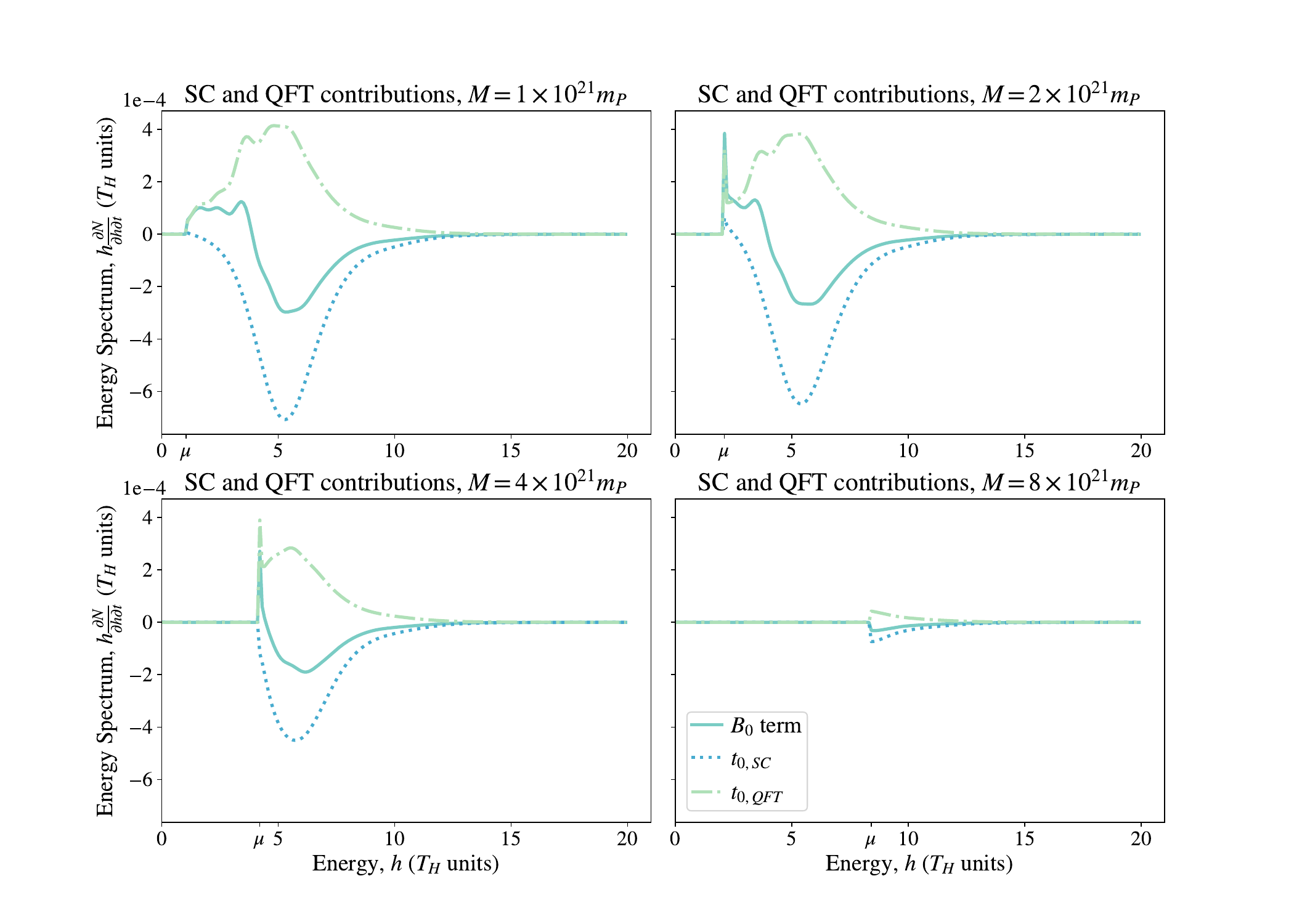}
    \caption{This figure shows the electron energy spectrum contribution from the $B_0$ term in the solid blue-green line, as well as the contributions from its semi-classical (blue dotted line) and QFT (green dash-dot line) constituent terms for four different black hole masses.} 
    \label{fig:SC_QFT}
\end{figure}

For both $B_0$ and $B_1$ corrections, except near $h=\mu$, there are only two terms ($t_{01}$ and $t_{03}$ for $B_0$, and $t_{11}$ and $t_{12}$ for $B_1$) providing significant contributions. In both cases there is one term that provides a positive correction ($t_{01}$ and $t_{11}$) and one that provides a negative correction ($t_{03}$ and $t_{12}$), and in both cases the relative contributions of these two terms oscillate as a function of energy.

When Eq.~(\ref{eq:B0_sc}) and Eq.~(\ref{eq:B0_QFTterms}) are multiplied by $(2j+1)/(2\pi)$, summed over $k$ and rewritten in terms of $t_{0i}$, the semi-classical correction and the full QFT contribution are given by
\begin{align*}
    t_{0,sc} \equiv - \sum_k \left(\frac{2j+1}{2\pi}\right) \frac{\alpha}{M} B_0(k,h) \vert_{\rm sc} &= t_{01}+2t_{03}\displaybreak[1]\\
    t_{0,QFT} \equiv - \sum_k \left(\frac{2j+1}{2\pi}\right) \frac{\alpha}{M} B_0(k,h) \vert_{\rm QFT} &= t_{02}-t_{03}+t_{04}+t_{05}+t_{06}+t_{07}+t_{08}+t_{09}.
\end{align*}
The semi-classical term $t_{0,sc}$ will have a larger magnitude compared to $t_{0,QFT}$ in cases where $t_{01}$ and $t_{03}$ dominate, which occurs everywhere except near the $h=\mu$ limit (where terms such as $t_{08}$ and $t_{09}$ contribute significantly). However, the negative $t_{03}$ contribution in $t_{0,QFT}$ implies that the QFT terms are non-negligible wherever the semi-classical terms are significant, and will reduce the magnitude of $B_0$ compared to the semi-classical prediction. 

This effect is noticeable in Fig. \ref{fig:SC_QFT} where $t_{0,sc}$  has a similar (but larger) magnitude and opposite sign compared to $t_{0,QFT}$ for $h>>\mu$. However, near $h=\mu$ for all but the highest mass, the QFT terms are dominant. The contribution of the QFT terms to $B_0$ in the $h=\mu$ limit is larger for smaller masses, and is largest in the case of $M=1\times10^{21} \ m_{\rm Pl}$ where $t_{0,sc}$ is negligible. 

\subsubsection{Verification of the Limit of $B_1$ for Large Black Hole Masses}\label{subsubsec:B1 sign}

\begin{figure}
    \centering
    \includegraphics[width=0.75\linewidth]{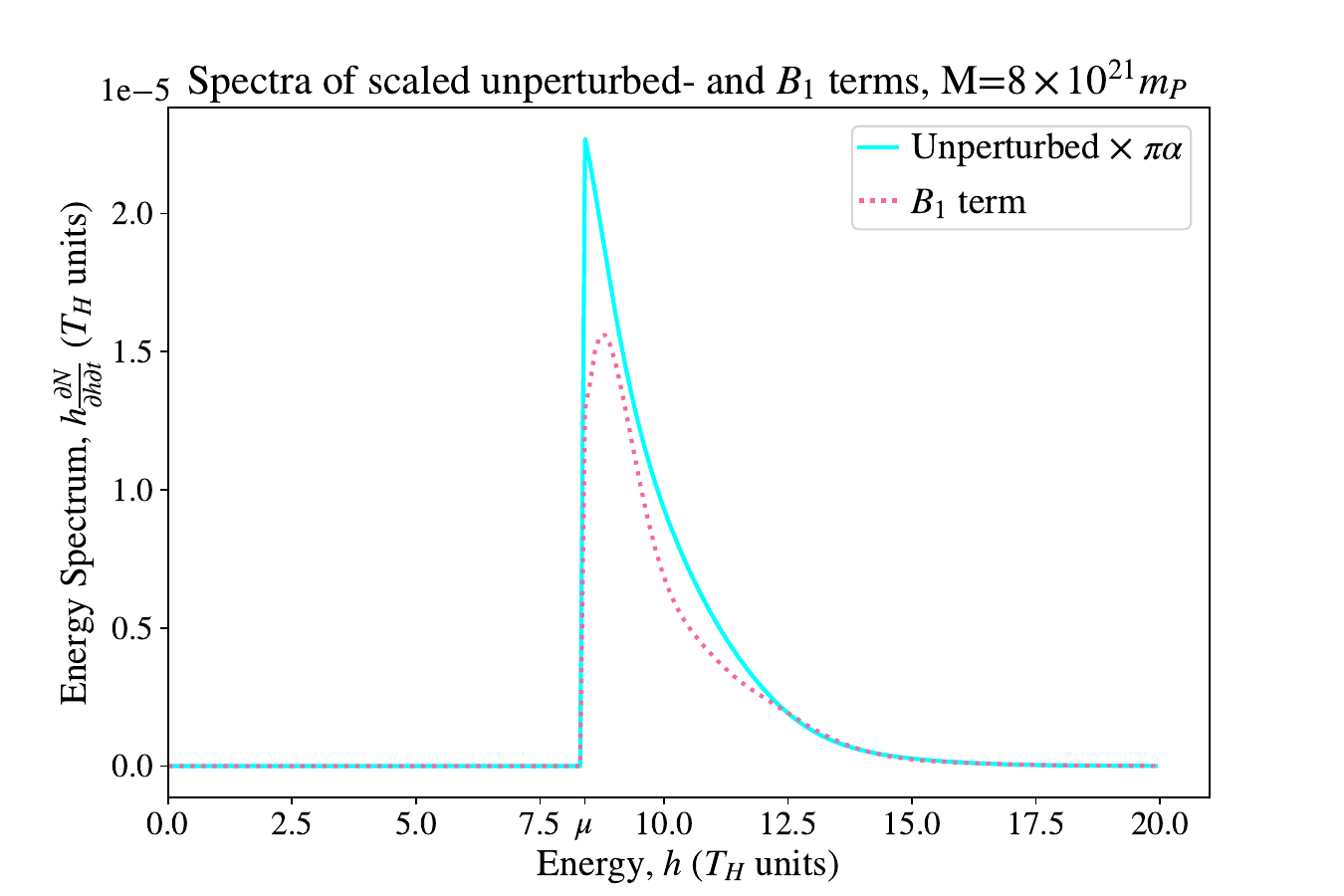}
    \caption{This figure shows the total contribution of the $B_1$ term to the $\mathcal{O}(\alpha)$ correction in the dotted pink line for a black hole mass of $8\times10^{21}$ Planck masses. For comparison, we show the energy spectrum of the free field emission scaled by $\pi \alpha$ (see Eq.~\ref{eq:pialpha}) in the solid cyan line.} 
    \label{fig:B1sign}
\end{figure}

When introducing the stochastic charge effect, we suppose that the particle number emission per unit time ($\dot{N}_{\pm}$) in the high mass, low Hawking temperature limit ($T_H<<\mu$) is
\begin{equation}
    \dot{N}_{\pm}=C e^{\pm 4 \pi \alpha Z}
\end{equation}
where $C$ is some proportionality constant. Upon promoting $Z$ to an operator $\hat{\cal Z}$, we can determine the expectation value of the particle number emission per unit time through Taylor expansion. Since $\langle \hat{\cal Z} \rangle=0$, this simplifies to
\begin{equation}
    \langle \dot{N}_{\pm} \rangle = C (1+8\pi^2 \alpha^2 \langle \hat{\cal Z}^2 \rangle+...) = C (1+\pi \alpha+...)
\label{eq:pialpha}
\end{equation}
as $\langle \hat{\cal Z}^2 \rangle \rightarrow \frac{1}{8 \pi \alpha}$ in this limit \cite{PhysRevD.16.2402}. Because the correction term proportional to $\langle \hat{\cal Z}^2 \rangle$ is $B_1$, we expect the $B_1$ correction to the particle number emission rate to be $\pi \alpha$ times the free field emission in the limit $T_H<<\mu$. 

Figure \ref{fig:B1sign} shows the total contribution of all $B_1$ terms to the energy spectrum as well as the free field energy spectrum scaled by $\pi \alpha$ for the highest mass for which we computed the numerical results. Since the energy spectrum is the number spectrum scaled by the energy, we expect that in the $T_H<<\mu$ limit, the energy spectrum of these two terms should be very similar. The curves have similar shape, with the scaled unperturbed term peaking at $h=\mu$ and the $B_1$ term only peaking $0.306T_H$ away. The largest relative difference between the two curves is $0.43$, occurring at $h=8.41T_H$, which is the lowest $h$ value greater than $\mu$ for which the numerical calculation was performed. Both curves are positive on the calculated range and both tend towards zero in the high energy limit.

\subsubsection{Electron Spectrum Corrections}\label{subsubsec:full results}

\begin{figure}
    \centering
    \includegraphics[width=\linewidth]{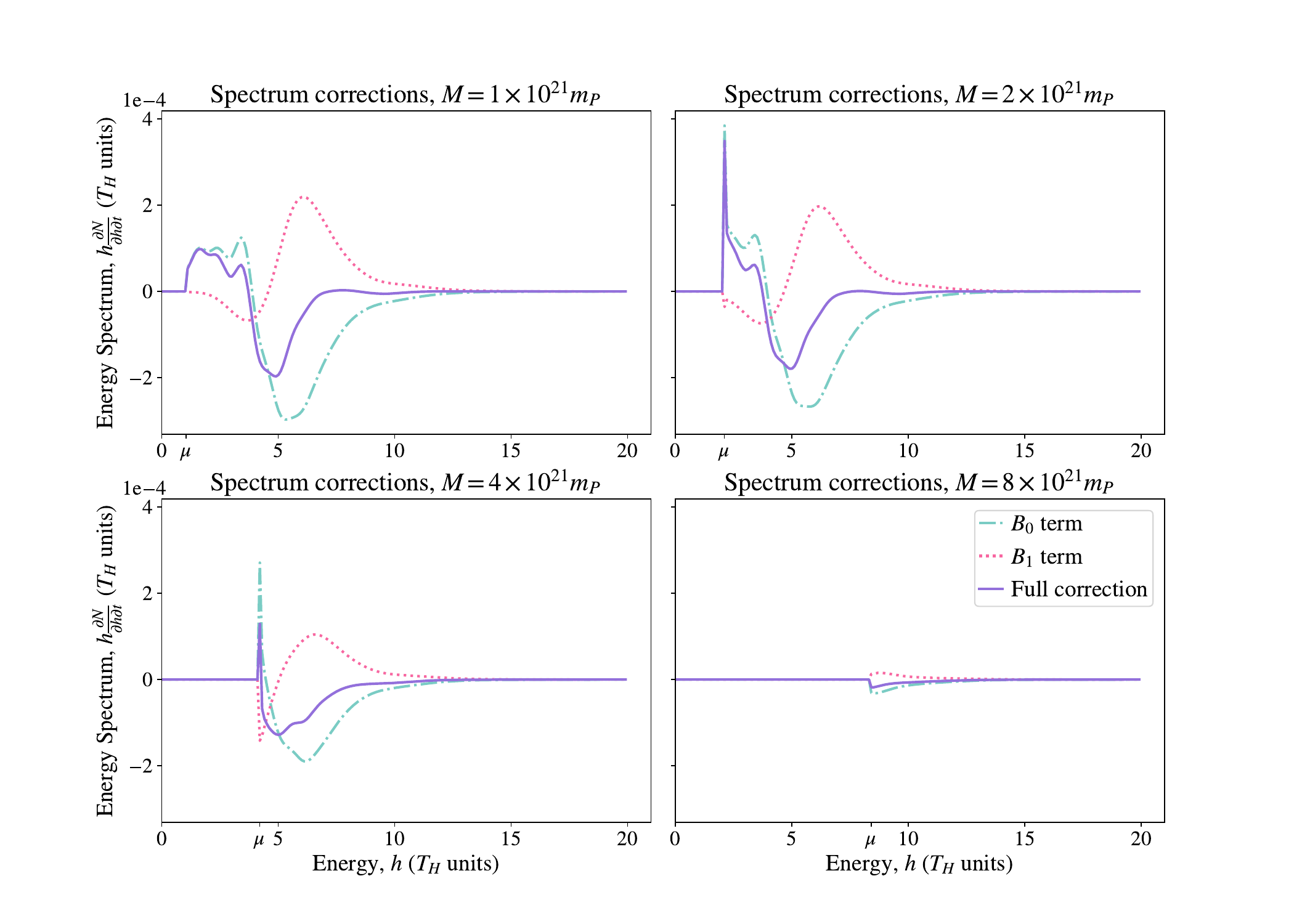}
    \caption{This figure shows the contributions to the stochastic charge $\mathcal{O}(\alpha)$ electron energy spectrum correction from $B_0$ terms (blue dash-dot line) and $B_1$ terms (pink dotted line) as well as the sum of these which make up the total correction (solid purple line) for four different black hole masses.}
    \label{fig:4panelB0B1breakdown}
\end{figure}

The contributions from $B_0$ and $B_1$ to the full stochastic charge $\mathcal{O}(\alpha)$ correction to the electron energy spectrum are of similar magnitude. Fig. \ref{fig:4panelB0B1breakdown} shows that for the three lower masses (top two panels and bottom left panel), the $B_0$ correction to the energy spectrum is positive at low energies (near $h=\mu$), negative at intermediate energies and tending toward 0 for high energies. For the highest mass (bottom right panel) it is negative at low energies and also tends toward 0 at high energies. Conversely, it shows that the $B_1$ correction is negative at low energies and positive at intermediate energies for the three lower masses, and positive at low energies for the highest mass. The $B_1$ correction also tends towards 0 in the high energy limit for all masses. The full correction therefore contains a complex interplay between the two terms. Neither term can be reasonably neglected in the energy range where the $\mathcal{O}(\alpha)$ correction is significant, except in the low energy limit where $B_0$ is dominant. As shown in Fig. \ref{fig:SC_QFT}, this limit is where QFT terms are dominant in $B_0$ for lower masses, and are therefore the most significant contributor to the full correction in this limit.

\begin{figure}
    \centering
    \includegraphics[width=\linewidth]{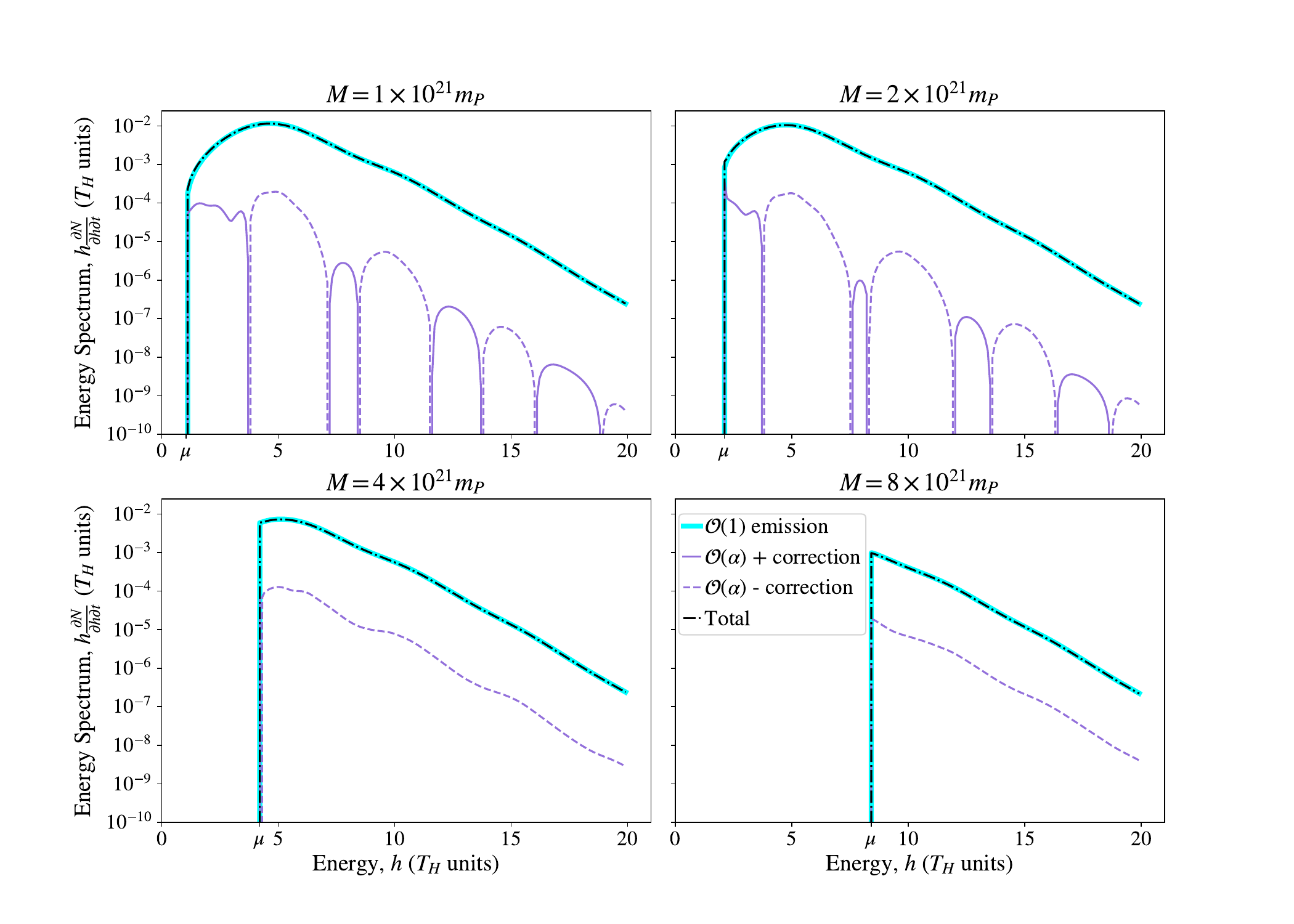}
    \caption{This figure shows the unperturbed electron energy spectrum in cyan, total stochastic charge $\mathcal{O}(\alpha)$ correction to the electron energy spectrum in purple (solid lines indicating a positive correction and dashed lines indicating a negative correction), and the corrected electron energy spectrum (the sum of the unperturbed and correction spectra) in the black dash-dot line for four different black hole masses. The corrections are approximately two orders of magnitude less than the unperturbed spectra.}
    \label{fig:4panelspectrum}
\end{figure}

The unperturbed $\mathcal{O}(1)$ electron energy spectrum, the $\mathcal{O}(\alpha)$ correction and their sum are shown in Fig. \ref{fig:4panelspectrum} for the four different black hole masses. In all four cases, the $\mathcal{O}(\alpha)$ correction is approximately 2 orders of magnitude smaller than the unperturbed spectrum, meaning it provides a percent-level contribution to the total spectrum. For the two lower masses (top two panels), the $\mathcal{O}(\alpha)$ correction oscillates as a function of energy between positive and negative contributions, and for the two higher masses (bottom two panels) the correction is negative across the computed energy range.
The oscillation is associated with the quantization of angular momentum: there is one ``cycle'' of oscillation for each partial wave ($j=\frac12, \frac32, \frac52...$) that can penetrate the angular momentum barrier as the energy is increased. 

The correction to the number and energy emission rates due to the stochastic charge effect is shown in Table~\ref{tab:emission_rates} for four different black hole masses. The overall changes to the emission rate are relatively modest, with the largest correction being a $1.75\%$ decrease at the largest mass considered. This order of magnitude is unsurprising for a process that is ${\mathcal O}(\alpha)$. We note that the semiclassical calculation of \citet{PhysRevD.16.2402} --- which is included in our calculations --- gave larger corrections, with decreases in both number and emission rates reaching $\sim 5\%$ at $M\mu \sim 0.2$ or $M\approx 5\times 10^{21}m_{\rm Pl}$. Our smaller net correction is at least partly a result of the aforementioned cancellation of contributions with positive and negative signs. We note, however, that over most of the parameter space we also find a net decrease in the charged lepton emission rate. 

\def\arraystretch{1.5}
\begin{table}
\begin{tabular}{ ccc|cc|cc } 
\hline
\multicolumn{3}{c|}{Mass} & \multicolumn4c{Corrections} \\
\hline
 & & & \multicolumn{2}{c|}{Number Emission Rate $dN_{e^-}/dt$} & \multicolumn2c{Energy Emission Rate $dE_{e^-}/dt$} \\ 
 \cline{4-7}
 $m_{\rm Pl}$ & & g  & Absolute & Relative & Absolute & Relative \\
\hline
$1\times10^{21}$ & & $2.18\times 10^{16}$ & $+1.88\times10^{-5}T_{\rm H}$  & $+0.18\%$ & $-1.69\times10^{-4}T_{\rm H}^2$ & $-0.37\%$ \\ 
$2\times10^{21}$ & & $4.35\times 10^{16}$ & $-8.47\times10^{-6}T_{\rm H}$ & $-0.10\%$ & $-1.86\times10^{-4}T_{\rm H}^2$ & $-0.46\%$ \\ 
$4\times10^{21}$ & & $8.71\times 10^{16}$ & $-5.31\times10^{-5}T_{\rm H}$ & $-1.36\%$ & $-3.12\times10^{-4}T_{\rm H}^2$ & $-1.36\%$ \\ 
$8\times10^{21}$ & & $1.74\times 10^{17}$ & $-3.08\times10^{-6}T_{\rm H}$ &  $-1.75\%$ & $-3.01\times10^{-5}T_{\rm H}^2$ & $-1.75\%$ \\ 
\hline
\end{tabular}
\caption{This table indicates the $\mathcal{O}(\alpha)$ number and energy emission rate corrections for four different black hole masses. ``Absolute'' corrections in the table are for electrons only; these should be doubled if the total (electrons + positrons) is desired.}
\label{tab:emission_rates}
\end{table}


\section{Discussion}

A comprehensive modeling of the Hawking emission spectra from PBHs in the $M\sim {\rm few}\times 10^{16}\,{\rm g}$ mass range is key to testing them as a dark matter candidate. 
This paper is part of a series calculating the Hawking emission spectra for photons, electrons, and positrons to $\mathcal{O}(\alpha)$ using a perturbative QED approach on a Schwarzschild spacetime. In this paper, we consider the ``long range monopole'' (0L) part of the interaction, and break the result down into two pieces: a ``$B_1$'' term associated with the infrared-divergent charge-squared operator $\langle \hat{\cal Z}^2 \rangle$ operator, and a ``$B_0$'' term that does not contain $\langle \hat{\cal Z}^2 \rangle$. We show that the ``stochastic charge'' effect identified from semi-classical arguments by \citet{PhysRevD.16.2402} --- and interpreted there as the variance of the charge of the black hole, $\langle Z^2 \rangle$ --- arises in QED, thus confirming the conjecture (``Conjecture \#2'') made in \citet{2024arXiv240709724V}. The interpretation of the term is different, however: rather than involving a fluctuating charge of the black hole, {\bfseries all the interactions that lead to the stochastic charge in the QED picture happen outside the horizon}. The black hole becomes surrounded by a plasma of $e^\pm$ pairs (as described in the Boulware frame), and it is this cloud of plasma whose charge is stochastically fluctuating.

We now turn our attention to $B_0(k,h)$, given by Eq.~(\ref{eq:B_0_final}), and discuss the origin of the terms contained within this function. We separate terms that appear in the semi-classical emission spectra into a new function $B_0(k,h) \vert_{\rm sc}$, given by Eq.~(\ref{eq:B0_sc}), while terms that only appear in the context of our QFT approach are contained within $B_0(k,h) \vert_{\rm QFT}$, given by Eq.~(\ref{eq:B0_QFTterms}). In the semi-classical perspective, the emission of an electron or positron will cause the black hole to acquire an equal and opposite charge, giving rise to a Coulomb potential at the horizon that will perturb the outgoing electron and positron emission rate. The QFT calculation introduces two new components. First, there are some terms associated with the electron self-energy; we defer a detailed discussion of these to future work because the long-range monopole term is one of several pieces of the electron self-energy. Second, there are exchange terms from fermion-fermion scattering outside the horizon. In addition, there are terms that arise from the resolution of the singularities in these terms. We evaluated all these effects in $B_0(k,h)\vert_{\rm QFT}$ and found each of them to be finite and well-behaved. We conclude that all divergences in the long-range monopole electrostatic sector, i.e., from $H_{\rm int, \Phi, 0L}$, cancel. This is a formal result, but very important to the success of the full calculation.

Our numerical results indicate that the stochastic charge effect provides a percent-level contribution to the $e^\pm$ energy spectrum. The QFT terms $t_{0,QFT}$ are the dominant contributor to this correction in the low energy limit for lower black hole masses. While the semi-classical terms $B_1$ and $t_{0,sc}$ make up the majority of the correction at higher energies, the QFT terms remain non-negligible. Percent-level energy spectrum accuracy at energies of a few Hawking temperatures higher than $h=\mu$ can be obtained by including only $t_{01}$, $t_{03}$, $t_{11}$ and $t_{12}$ terms. Since $t_{01}$ and $t_{11}$ are opposite in sign compared to $t_{03}$ and $t_{12}$, neglecting any of these terms may lead to an overestimate of the ${\mathcal O}(\alpha)$ stochastic charge effect correction in this limit.


There are still several more steps to complete the computation of the electron spectrum at ${\mathcal O}(\alpha)$, as shown in the flow chart of \citet{2024arXiv240709724V}. The corrections divide into two parts: the electrostatic sector (mediated by $\Phi$ in Coulomb gauge) and the electrodynamic sector (mediated by $A_i$). This paper has considered the long-range monopole correction from the electrostatic sector Hamiltonian $H_{\rm int, \Phi, 0L}$. While this encapsulates the infrared-divergent piece, the remaining electrostatic terms (the short-range monopole Hamiltonian $H_{\rm int, \Phi, 0S}$, as well as $\ell \neq 0$ which are all short range) still need to be evaluated. Also, the dissipative and conservative terms in the electrodynamic sector need to be evaluated (just as they do for photons; see \citet{Koivu_2025} for our treatment of the dissipative corrections to the photon spectrum). These calculations are in progress and will be presented in future articles.

The success of the semi-classical approach with QED suggests that similar approaches might be pursued for the other gauge forces in the Standard Model that become relevant at higher energies (i.e., smaller black holes). Such smaller PBHs would not constitute a significant fraction of dark matter, but could be a ``tail'' of low-mass objects that evaporated in the early Universe, or (if $M\sim 5\times 10^{14}\,$g) are evaporating today \cite{1991PhRvD..44..376M}. At the lowest masses ($M \lesssim 10^{11}\,{\rm g}$ or $T \gtrsim 100\,$GeV), one expects the Standard Model ``charge'' of the black hole (including the plasma surrounding the event horizon) to undergo a random walk in the space of ${\rm SU}(3)_{\rm QCD}\times {\rm SU}(2)_{\rm weak}\times {\rm U}(1)_Y$ representations. The charge variance (or even stability) of this random walk for the non-Abelian gauge groups has not been investigated. Finally, we note that the ${\rm U}(1)_Y$ coupling constant (where $Y$ is weak hypercharge) increases with energy, so even the ``order $\alpha$'' corrections may become large in the final stages of PBH evaporation --- although this picture could be further complicated by inclusion of beyond-Standard Model particles \cite{2020PhRvD.101e5006B}, even if they are ${\rm SU}(3)_{\rm QCD}\times {\rm SU}(2)_{\rm weak}\times {\rm U}(1)_Y$ scalars \cite{2026PhRvD.113d4038E}, or in the context of new gauge groups or extra dimensions \cite{2022PhRvD.105j3508F}.


\appendix

\section{Interaction integrals, derivatives, and pole structure}
\label{app:E-Lambda}
We are now interested in how ${\cal A}_{XX'k}(h,h')$ and its derivatives at $h'=h$ relate to the singular behavior when poles are displaced by $\epsilon$. A subtle issue arises when calculating integrals that have two poles of the form $(h-h'\pm i\epsilon)^{-1}$. The integrand is of order $\epsilon^{-2}$ over a range $\Delta h'$, which is also of order $\epsilon$, and thus the integral is naively of order $1/\epsilon$. Therefore, it is required to determine the next-to-leading order (in $\epsilon$) terms in ${\cal I}^{(1)}_{XX'k}(h,h')$ which need to be taken into consideration. To account for this, we introduce an additional $\epsilon D_{XX'k}(h)$ term that will disappear in the $\epsilon \rightarrow 0$ limit unless there is an additional $\mathcal{O}(1/\epsilon)$ term that is not being accounted for. Likewise, we introduce an additional $\epsilon^2 N_{XX'k}(h)$ term to account for any additional $\mathcal{O}(1/\epsilon^2)$ terms that are unaccounted for. Therefore, our new definition for ${\cal I}^{(1)}_{XX'k}(h,h')$ is given by
\begin{equation}
    {\cal I}^{(1)}_{XX'k}(h,h') = \frac{{\cal A}_{XX'k}(h,h') + \epsilon D_{XX'k}(h) - \frac12 i\epsilon^2 N_{XX'k}(h)}{h-h'-i\epsilon} + 2\pi \delta_\epsilon(h-h') \delta_{X,up} \delta_{X', up},
\label{eq:F1.}
\end{equation}
where $\delta_\epsilon(h-h')$ is given by Eq.~(C11) in \citet{2024arXiv240709724V}. We have taken the modified pole to order $\epsilon^2$ (the highest order that we need in this paper) and note that ${\bf D}$ and ${\bf N}$ are functions of $h$, not $h'$; and that ${\bf D}$ is anti-Hermitian while ${\bf N}$ is Hermitian. (We package objects with $XX'$ indices into $2\times 2$ matrices with a boldface symbol.) We use the explicit definition of ${\cal A}_{XX'k}(h,h')$ given in \citet{2024arXiv240709724V}:
\begin{equation}
    {\cal A}_{XX'k}(h,h') = (h-h') \underline{{\cal I}}^{(1)}_{XX'k}(h,h')  = (h-h') \lim_{\epsilon \rightarrow 0} {\cal I}^{(1)}_{XX'k}(h,h' \vert \epsilon)  ,
\label{eq:A-def}
\end{equation}
where we recall that ${\cal I}^{(1)}_{XX'k}(h,h')$ has a singularity at $h'=h-i\epsilon$. Defining $S_{\frac12 kh} \equiv R^\ast_{\frac12 kh} e^{2i \arg T_{1/2, k,h}}$, we list the components of Eq.~(\ref{eq:A-def}) below:
\begin{eqnarray}
    {\cal A}_{\rm in,in,k}(h,h') &\equiv (h-h') \underline{{\cal I}}^{(1)}_{\rm in,in,k}(h,h') &= -i T^\ast_{ \frac12,k,h}T_{ \frac12,k,h'}  + (h-h') \Lambda_{\rm in,in,k}(h,h'),
\nonumber \\
    {\cal A}_{\rm in,up,k}(h,h') &\equiv (h-h') \underline{{\cal I}}^{(1)}_{\rm in,up,k}(h,h') &= i T^\ast_{ \frac12 kh} S_{\frac12 kh'} + (h-h') \Lambda_{\rm in,up,k}(h,h'),
\nonumber \\
    {\cal A}_{\rm up,in,k}(h,h') &\equiv (h-h') \underline{{\cal I}}^{(1)}_{\rm up,in,k}(h,h') &= i S^\ast_{ \frac12 kh} T_{ \frac12 kh'} + (h-h') \Lambda_{\rm up,in,k}(h,h'),
\nonumber \\
    {\cal A}_{\rm up,up,k}(h,h') &\equiv (h-h') \underline{{\cal I}}^{(1)}_{\rm up,up,k}(h,h')  &= i (1 - S^\ast_{\frac12 kh} S_{ \frac12 kh'} ) + (h-h') \Lambda_{\rm up,up,k}(h,h').
\label{eq:A-table}
\end{eqnarray}
(see Appendix C of \citet{2024arXiv240709724V}). This expression also serves as the definition of the $\Lambda$-terms when we go to higher order in the singularities; note the exact Hermiticity ${\boldsymbol\Lambda}_k^\dagger(h,h') = {\boldsymbol\Lambda}_k(h',h)$ for real $h$ and $h'$.

In the $h=h'$ limit, the ${\cal A}$-terms reduce to simple expressions:
\begin{align}
{\cal A}_{{\rm in,in},k}(h,h) =& -i|T_{\frac12kh}|^2,
~~~~~~~~~~~~~
&{\cal A}_{{\rm in,up},k}(h,h) = iR_{\frac12kh}^\ast T_{\frac12kh},
\nonumber \\
{\cal A}_{{\rm up,in},k}(h,h) =& iT_{\frac12kh}^\ast R_{\frac12kh},
~~~~~~~~~~
&{\cal A}_{{\rm up,up},k}(h,h) = i|T_{\frac12kh}|^2.~~~~
\label{eq:A-val}
\end{align}
The calculation of the electron emission spectra will require us to determine the $1^{\text{st}}$ and $2^{\text{nd}}$ derivatives of Eq.~(\ref{eq:A-table}) evaluated at $h'=h$. We define these derivatives as
\begin{equation}
\partial_{h'} {\cal A}_{XX'k}(h,h')\vert_{h'=h} \equiv {\cal E}_{XX'k}(h) ~~~ {\rm and}~~~
\partial_{h'}^2 {\cal A}_{XX'k}(h,h')\vert_{h'=h} \equiv Q_{XX'k}(h).
\label{eq:aeq}
\end{equation}
We may write the four components of ${\cal E}_{XX'k}(h)$ in terms of transmission and reflection coefficients:
\begin{eqnarray}
    {\cal E}_{{\rm in, in},k} (h) &=& -i T^\ast_{\frac12 kh} \frac{dT_{\frac12 kh}}{dh} - \Lambda_{\rm in,in,k}(h,h), \nonumber\\
     {\cal E}_{{\rm in, up},k} (h) &=& i T_{\frac12 kh} \frac{dR^\ast_{\frac12 kh}}{dh} - 2R^\ast_{\frac12 kh} T_{\frac12 kh} \frac{d}{dh} \arg T_{\frac12 kh} - \Lambda_{\rm in,up,k}(h,h), \nonumber\\
     {\cal E}_{{\rm up, in},k} (h) &=& i R_{\frac12 kh} e^{-2i\arg T_{\frac12 kh}} \frac{dT_{\frac12 kh}}{dh} - \Lambda_{\rm up,in,k}(h,h),~~{\rm and} \nonumber\\
     {\cal E}_{{\rm up, up},k} (h) &=& -i R_{\frac12 kh} \frac{dR^\ast_{\frac12 kh}}{dh} + 2|R_{\frac12 kh}|^2 \frac{d}{dh} \arg T_{\frac12 kh} - \Lambda_{\rm up,up,k}(h,h).
\label{eq:E2}
\end{eqnarray}
We can also express these derivatives in terms of the transmission probability ${\mathbb T}_{\frac12 kh} = |T_{\frac12 kh}|^2$:
\begin{eqnarray}
{\cal E}_{{\rm in,in},k}(h) &=& -\Lambda_{{\rm in,in},k}(h,h)
- \frac12i\frac{d{\mathbb T}_{\frac12kh}}{dh}
+ |T_{\frac12kh}|^2 \frac d{dh}\arg T_{\frac12kh},
\nonumber \\
{\cal E}_{{\rm in,up},k}(h) &=& -\Lambda_{{\rm in,up},k}(h,h)
- \frac12i \frac{T_{\frac12kh}}{R_{\frac12kh}} \frac{d{\mathbb T}_{\frac12kh}}{dh}
+R^\ast_{\frac12kh} T_{\frac12kh} \frac d{dh}\arg R_{\frac12kh}
-2 R^\ast_{\frac12kh}T_{\frac12kh} \frac d{dh}\arg T_{\frac12kh},
\nonumber \\
{\cal E}_{{\rm up,in},k}(h) &=& -\Lambda_{{\rm up,in},k}(h,h) + \frac12 i \frac{R_{\frac12kh}}{T_{\frac12kh}}\frac{d{\mathbb T}_{\frac12kh}}{dh}
-R_{\frac12kh}T_{\frac12kh}^\ast \frac{d}{dh}\arg T_{\frac12kh},~~{\rm and}
\nonumber \\
{\cal E}_{{\rm up,up},k}(h) &=& -\Lambda_{{\rm up,up},k}(h,h) + \frac12i \frac{d{\mathbb T}_{\frac12kh}}{dh}
- |R_{\frac12 kh}|^2 \frac{d}{dh} \arg R_{\frac12 kh} + 2|R_{\frac12 kh}|^2 \frac{d}{dh} \arg T_{\frac12 kh}.
\label{eq:E-SUB}
\end{eqnarray}
While defining the $2^{\text{nd}}$ derivative of ${\cal A}_{XX'k}(h,h')$ is crucial towards calculating how the semi-classical derivative $\partial^2_Z {\mathbb T}_{\frac12 kh}(0)$ relates to our QFT formalism (see Appendix~\ref{app:2deriveT} for a full derivation), we do not need the similar expressions for $Q_{XX'k}(h)$, so we will not list them here.

Having defined the components of the $\cal A$-terms and their derivatives, evaluated at $h'=h$, we wish to understand the mathematical structure of our corrections to Eq.~(\ref{eq:F1.}). The corrections to Eq.~({\ref{eq:F1.}}) are represented by the matrices ${\bf D}_k(h)$ and ${\bf N}_k(h)$. To determine what our corrections are, we wish to calculate the residue from our original definition of ${\cal I}^{(1)}_{XX'k}(h,h')$, given by 
\begin{equation}
    {\cal I}^{(1)}_{XX'k}(h,h') = \frac{{\ C}_{XX'k}(h,h')}{h-h'-i\epsilon} + 2\pi \delta_\epsilon(h-h') \delta_{X,up} \delta_{X', up} + \Lambda_{XX'k}(h,h')
\label{eq:F2.}
\end{equation}
where
\begin{align}
{C}_{{\rm in,in},k}(h,h) =& -iT^\ast_{\frac12kh}T_{\frac12kh'},
~&~~~~~~~~~~~~
{C}_{{\rm in,up},k}(h,h) = iT_{\frac12kh}^\ast S_{\frac12kh'},~~~~~~~\,
\nonumber \\
{C}_{{\rm up,in},k}(h,h) =& iS_{\frac12kh}^\ast T_{\frac12,kh'},
~&{\rm and}~~~~~~~~~~~
{C}_{{\rm up,up},k}(h,h) = i(1-S_{\frac12kh}S_{\frac12kh'})
\label{eq:C-val}
\end{align}
are the components of the matrix ${\bf C}_k(h,h')$, at $h'=h-i\epsilon$ and Taylor expand this result around $h'=h$. Performing the same procedure for Eq.~(\ref{eq:F1.}), we can compare terms to derive the explicit definition of $D_{XX'k}(h)$.
Calculating the residue of Eqs.~(\ref{eq:F1.}) and (\ref{eq:F2.}) evaluated at $h'=h-i\epsilon$ and Taylor expanding around $h'=h$, we get:
\begin{equation}
i{\mathbb I} + {\bf A}_k(h,h) + \epsilon [{\bf D}_k(h)-i{\bf E}_k(h)] + \frac{\epsilon^2}2[-i{\bf N}_k(h)-{\bf Q}_k(h)] + ...
= {\bf C}_k(h,h) -i\epsilon \partial_{h'}{\bf C}_k(h,h')|_{h'}
- \frac{\epsilon^2}2 \partial_{h'}{\bf C}_k(h,h')|_{h'} + ...,
\end{equation}
where ${\mathbb I}$ is the $2\times 2$ identity matrix and the functions $S_{\frac12 kh}$, $T_{\frac12 kh}$, and ${\cal A}_{XX'k}(h,h')$ are analytically continued as functions of $h'$ to accept complex energies.

Equating the $\mathcal{O}(\epsilon)$ terms, and using Eq.~(\ref{eq:E2}), there is a full cancellation except for $\Lambda$-type terms. Thus, 
\begin{equation}
    D_{XX'k}(h) = - i \Lambda_{XX'k}(h,h)
\end{equation}
which explains the form of Eq.~(\ref{eq:I(1)-modified}). Equating $\mathcal{O}(\epsilon^2)$ terms, the result for ${\bf N}$ and ${\bf Q}$ is similar:
\begin{equation}
{\bf N}_k(h) - i{\bf Q}_k(h) = \left( \begin{array}c -T^\ast_{\frac12kh} \\ S^\ast_{\frac12kh} \end{array} \right)
\left( \begin{array}{cc}
\partial_h^2T_{\frac12kh} & -\partial_h^2S_h
\end{array} \right).
\label{eq:NQ.1}
\end{equation}
These results, combined with identity $-R_{\frac12kh}T^\ast_{\frac12kh} + T_{\frac12kh}S^\ast_{\frac12kh}= 0$, prove the following convenient lemmas:
\begin{equation}
\sum_X w_X [{\cal E}_{XX'k}(h) + \Lambda_{XX'k}(h)] = 0
~~{\rm and}~~
\sum_X w_X [N_{XX'k}(h) - iQ_{XX'k}(h)] = 0.
\label{eq:w0}
\end{equation}

\section{Singular integrals}
\label{app:singular}

This appendix derives singular integrals that are used elsewhere in this paper.

\subsection{Double pole integrals with one small parameter}
\label{eq:double}

We begin with the integrals of the form:
\begin{equation}
    \int_0^\infty \frac{dh'}{2\pi} \frac{2\pi\delta_\epsilon(h-h')}{h-h'+i\epsilon} \Phi(h') = i \int_0^\infty \frac{dh'}{2\pi} \left(\frac{1}{h-h'+i\epsilon} - \frac{1}{h-h'-i\epsilon} \right) \frac{\Phi(h')}{h-h'+i\epsilon},
\end{equation}
where we substituted Eq.~(C11) from \citet{2024arXiv240709724V} for $\delta_\epsilon(h-h')$ which resembles that of a Dirac delta function in the limit where $\epsilon$ is small.
We make the substitution 
$x = h'-h = \epsilon \tan q$, to arrive at:
\begin{equation}
\int_0^\infty \frac{dh'}{2\pi} \frac{2\pi\delta_\epsilon(h-h')}{h-h'+i\epsilon} \Phi(h') =
\frac i{2\pi\epsilon} \int_{-\pi/2 + \tan^{-1}(\epsilon/h)}^{\pi/2} \Phi(h + \epsilon\tan q)\,(1+e^{2iq})\,dq.
\end{equation}
We Taylor expand $\Phi$ around $h$ as $\Phi(h) + \epsilon \tan q\,\Phi'(h)$ (in the limit of $\epsilon \rightarrow 0^+$, we may drop the higher-order terms), where the $'$ on $\Phi$ is a derivative with respect to $h$.
There is a technicality at the lower limit of the integral: in principle, we should not set the lower limit to $-\pi/2$. However, if $\Phi$ is bounded, the integrand goes to zero as $q\rightarrow -\pi/2$ (since $1+e^{2iq}\rightarrow 0$),  so the error in the integral from setting the lower limit to $-\pi/2$ is higher order in $\epsilon$. Then
\begin{eqnarray}
    \int_0^\infty \frac{dh'}{2\pi} \frac{2\pi\delta_\epsilon(h-h')}{h-h'+i\epsilon} \Phi(h') &=&
\frac i{2\pi\epsilon} \left[ \Phi(h) \int_{-\pi/2}^{\pi/2} (1+e^{2iq})\,dq
+ \epsilon \Phi'(h) \int_{-\pi/2}^{\pi/2} \tan q\, (1+e^{2iq})\,dq
+ ...
\right]    
\nonumber \\ &=&
    - \frac{i}{2\epsilon} \Phi(h) - \frac{1}{2} \Phi'(h) + \mathcal{O}(\epsilon).
\label{eq:int1}
\end{eqnarray}
Similarly:
\begin{equation}
\int_0^\infty \frac{dh'}{2\pi} \frac{2\pi\delta_\epsilon(h-h')}{h-h'-i\epsilon} \Phi(h')
= \frac{i}{2\epsilon} \Phi(h) - \frac{1}{2} \Phi'(h) + \mathcal{O}(\epsilon).
\label{eq:-i}
\end{equation}
We will also encounter integrals of the following form (which will contain a principal part):
\begin{equation}
\int_{h-\gamma}^{h+\gamma} \frac{dh'}{2\pi} \frac{\Phi(h')}{(h-h'-i\epsilon)(h-h'+i\epsilon)}
= \frac{\Phi(h)}{2\epsilon} + {\cal O}(\epsilon/\gamma, \gamma),
\end{equation}
where we envision both that $\gamma$ is small (so that we can Taylor-expand $\Phi$ around $h$) and $\epsilon\ll \gamma$ (so that we can treat the singularity by the same substitution $h'=h+\epsilon\tan q$ and treat the limits of $q$ as $-\pi/2<q<\pi/2$). Since this integral is the difference between the full integral and the principal part, we have
\begin{equation}
\lim_{\epsilon\rightarrow 0^+} \int_0^\infty \frac{dh'}{2\pi} \frac{\Phi(h'|\epsilon)}{(h-h'-i\epsilon)(h-h'+i\epsilon)}
= {\mathbb P}\int_0^\infty \frac{dh'}{2\pi} \frac{\Phi(h'|0)}{(h-h')^2}
+ \lim_{\epsilon\rightarrow 0^+} \frac{\Phi(h|\epsilon)}{2\epsilon},
\label{eq:PPI}
\end{equation}
where ${\mathbb P}$ denotes a principal part. In some cases, we may be interested only in the real or the imaginary part of the integral, in which case it only matters that the appropriate limits for that part converge.

Finally, it is helpful to be able to express double pole integrals with the poles on the same side of the real axis in terms of principal parts. A similar approach yields
\begin{equation}
\lim_{\epsilon\rightarrow 0^+} \Im
\int_0^\infty \frac{dh'}{2\pi} \frac{\Phi(h')}{(h-h'+i\epsilon)^2} = 
{\mathbb P}
\int_0^\infty \frac{dh'}{2\pi} \frac{\Im \Phi(h')}{(h-h')^2} + \frac12\Re \Phi'(h)
~~{\rm if}~~ \Phi(h)~{\rm is~real}.
\label{eq:PPII}
\end{equation}

\subsection{Double and triple poles with two small parameters}

Here we address integrals that contain poles at $h'=h\pm i\epsilon$ and $h'=h+i\eta$ where the location of these poles on the complex plane and the order of which quantity approaches zero first matters. We are interested in taking the limit as $\epsilon\rightarrow 0^+$ first; and {\em then} we take the limit as $\eta\rightarrow 0^+$ (if $\eta$ appears in the problem) which is the physical limit.

In many cases, the form of $\Phi(h')$ will be
\begin{equation}
    \Phi(h') = \frac{\cYu(h')}{h-h'+i\eta},
\label{eq:Phi-form}
\end{equation}
where $\cYu(h')$ is analytic and continuous in the vicinity of $h'\approx h$. Noting that $\Phi(h) = -i\cYu(h)/\eta$, and its derivative is $\Phi'(h) = -i \cYu'(h)/\eta-\cYu(h)/\eta^2$, Eq.~(\ref{eq:int1}) reduces to
\begin{equation}
    \int_0^\infty \frac{dh'}{2\pi} \frac{2\pi\delta_\epsilon(h-h')}{h-h'+i\epsilon} \frac{\cYu(h')}{h-h'+i\eta} = - \frac{\cYu(h)}{2\epsilon \eta} + \frac{\cYu(h)}{2\eta^2}+ \frac{i \cYu'(h)}{2\eta}  + \mathcal{O}(\epsilon).
\label{eq:defect1}
\end{equation}

We now introduce the ${\mathbb U}$ operator that moves all poles to the upper half complex plane: that is, it replaces all instances of $h-h'-i\epsilon$ in any expression $expr$ with $h-h'+i\epsilon$, and annihilates $\delta_\epsilon(h-h')$. This is the complement of the ${\mathbb V}$ operator, Eq.~(\ref{eq:V-expr}), in the sense that ${\mathbb U}[expr] + {\mathbb V}[expr] = [expr]$.
With this notation, Eq.~(\ref{eq:defect1}) implies:
\begin{equation}
{\mathbb V} \int_0^\infty \frac{dh'}{2\pi} \frac{2\pi\delta_\epsilon(h-h')}{h-h'+i\epsilon} \frac{\cYu(h')}{h-h'+i\eta} = - \frac{\cYu(h)}{2\epsilon \eta} + \frac{\cYu(h)}{2\eta^2}+ \frac{i \cYu'(h)}{2\eta}
+ \mathcal{O}(\epsilon)
\end{equation}
(trivially, since ${\mathbb U}$ annihilates this expression). We also see that
\begin{equation}
{\mathbb V} \int_0^\infty \frac{dh'}{2\pi} \frac{\cYu(h')}{(h-h'+i\eta)(h-h'+i\epsilon)(h-h'-i\epsilon)} = - \frac{i\cYu(h)}{2\epsilon \eta} + \frac{i\cYu(h)}{2\eta^2}- \frac{ \cYu'(h)}{2\eta}
+ \mathcal{O}(\epsilon),
\end{equation}
since the ${\mathbb V}$ operator acting on this integral gives $i$ times Eq.~(\ref{eq:defect1}).
Similar reasoning leads to
\begin{equation}
{\mathbb V} \int_0^\infty \frac{dh'}{2\pi} \frac{2\pi\delta_\epsilon(h-h')}{h-h'-i\epsilon} \frac{\cYu(h')}{h-h'+i\eta} = \frac{\cYu(h)}{2\epsilon \eta} + \frac{\cYu(h)}{2\eta^2}+ \frac{i \cYu'(h)}{2\eta}
+ \mathcal{O}(\epsilon).
\end{equation}
The fact that the $\delta_\epsilon$ function converges to the $\delta$-function implies:
\begin{equation}
{\mathbb V} \int_0^\infty \frac{dh'}{2\pi} \frac{2\pi\delta_\epsilon(h-h') \cYa(h')}{h-h'+i\eta} = 
-\frac{i\cYa(h)}\eta + \mathcal{O}(\epsilon),
\label{eq:ya1}
\end{equation}
and then
\begin{equation}
{\mathbb V} \int_0^\infty \frac{dh'}{2\pi} \frac{ \cYa(h')}{(h-h'+i\eta)(h-h'-i\epsilon)} = 
\frac{\cYa(h)}\eta + \mathcal{O}(\epsilon).
\label{eq:ya2}
\end{equation}

A final integral that we will want (and where the ${\mathbb V}$ operator acts trivially since the integral is annihilated by ${\mathbb U}$) is
\begin{equation}
{\mathbb V}\int_0^\infty \frac{dh'}{2\pi}
\,[2\pi\delta_\epsilon(h-h')]^2\,
\Phi(h') =  \frac2{\pi \epsilon} \int_{-\pi/2+\tan^{-1}(\epsilon/h)}^{\pi/2} \!\Phi(h+\epsilon\tan q)\,
\cos^2 q\,dq
=  \frac2{\pi \epsilon} \left[ \frac\pi 2 \Phi(h)
+ \mathcal O(\epsilon^2)\right]
= \frac1{\epsilon} \Phi(h)
+ \mathcal O(\epsilon).
\end{equation}
(In this case, it is important that $\cos^2 q$ goes to zero quadratically at $q=-\pi/2$, so that errors associated with the lower limit of the integral are higher-order than $\epsilon^2$.)
Using $\Phi$ of the form Eq.~(\ref{eq:Phi-form}) gives
\begin{equation}
{\mathbb V}\int_0^\infty \frac{dh'}{2\pi}
\,[2\pi\delta_\epsilon(h-h')]^2\frac{\cYu(h')}{h-h'+i\eta} =
\frac{-i\cYu(h)}{\epsilon\eta}
 + \mathcal O(\epsilon).
\label{eq:DE2}
\end{equation}

\subsection{Imaginary parts}

We now consider cases with functions $\cZh$ that are real for real arguments. Then we have the familiar rule
\begin{equation}
\lim_{\upsilon\rightarrow 0^+}
\Im \int_0^\infty \frac{dh'}{2\pi} \frac{\cZh(h')}{(h-h'+i\upsilon)} = -\frac12 \cZh(h).
\label{eq:Zh1}
\end{equation}
Integration by parts gives the sequence of rules:
\begin{equation}
\lim_{\upsilon\rightarrow 0^+}
\Im \int_0^\infty \frac{dh'}{2\pi} \frac{\cZh(h')}{(h-h'+i\upsilon)^2} = -\frac12 \cZh'(h)
~~~~{\rm and}~~~~
\lim_{\upsilon\rightarrow 0^+}
\Im \int_0^\infty \frac{dh'}{2\pi} \frac{\cZh(h')}{(h-h'+i\upsilon)^3} = -\frac14 \cZh''(h),
\label{eq:Zh3}
\end{equation}
where $\cZh'$ and $\cZh''$ are the first and second derivatives of $\cZh$ with respect to $h$. Note that all that matters in these cases is that the poles are located at $\Im h'$ infinitesimally positive --- it does not matter in this case whether they are $\epsilon$ or $\eta$ away from the real axis, or the order of the limits.

\section{Perturbations to the transmission coefficients}
\label{app:T2}

In the semi-classical stochastic charge picture, one considers how the single-particle electron transmission coefficients vary as a function of the charge $Z$ on the black hole, i.e., we have ${\mathbb T}_{\frac12kh}(Z)$. Since we want to make contact between the operator approach here and the semi-classical picture, we will try to expand this in terms of the interactions (${\cal A}$-functions). In this appendix, only one value of $k$ appears at a time, so this is suppressed for clarity.

The starting point is the complete set of single-particle radial states $|X,h)$, where $X\in\{{\rm in,up}\}$ and $h$ ranges over both positive and negative values (and with $|h|>\mu$ for $X=$ in). We use the $|...)$ notation for single-particle radial states to distinguish them from $|...\rangle$ describing the quantum state of the full system. As per \citet{PhysRevD.107.045004}, these satisfy the inner product
\begin{equation}
(X,h|X',h') = \int_{-\infty}^\infty
[
F_{Xkh}^\ast(r_\star) F_{X'k'h'}(r_\star) +
G_{Xkh}^\ast(r_\star) G_{X'k'h'}(r_\star)
]\,dr_\star = 4\pi |h| \delta_{XX'} \delta(h-h'),
\end{equation}
and consequently there is a resolution of the identity operator:
\begin{equation}
{\mathbb I} = \int_{-\infty}^\infty \frac{dh}{4\pi|h|} \sum_X |X,h)(X,h|,
\end{equation}
where the sum is only nonzero for quasi-bound states ($|h|<\mu$) when $X={\rm up}$ while $X={\rm in,up}$ for free electron states ($|h|>\mu$). For a full-length discussion on the explicit form of the radial electron wave functions, $F_{Xkh}(r_\ast)$ and $G_{Xkh}(r_\ast)$, in the $r_\ast \rightarrow \pm \infty$ limit, please refer to Eqs.~(39-42) of \citet{PhysRevD.107.045004}. These radial states are eigenstates of the single-particle unperturbed Hamiltonian:
\begin{equation}
H_0^{\rm SP} = \frac kr\sqrt{1-\frac{2M}r} \,{\boldsymbol\sigma}_1 +
i\partial_{r_\star}\,{\boldsymbol\sigma}_2 +
\mu\sqrt{1-\frac{2M}r}\,{\boldsymbol\sigma}_3.
\end{equation}
where $\boldsymbol{\sigma}_i$ ($i=1,2,3$) are the Pauli matrices. The inclusion of a charge on the black hole introduces a Coulomb-like perturbation to the single-particle Hamiltonian which will alter the transmission probability and the electron phase space density. In accordance with the discussion in \citet{2024arXiv240709724V}, we impose a cutoff $\propto e^{\epsilon r_\star}$ as one approaches the horizon, with the intention that we will take $\epsilon\rightarrow 0^+$. For definiteness, the form of this perturbation is given by
\begin{equation}
W^{\rm SP} = -\frac{Z\alpha}{r} e^{-\epsilon|r_\star|}.
\label{eq:C4}
\end{equation}

We introduce time-independent perturbation theory by starting the system with a perturbation $H^{\rm SP} = H_0^{\rm SP} + W^{\rm SP} e^{\eta t}$ (with a growing exponential such that the perturbation was turned off in the distant past; eventually we will take $\eta\rightarrow 0^+$). Then a single particle state can be expanded as a perturbation series in $W^{\rm SP}$:
\begin{equation}
|\varphi(t)) = |\varphi_0)e^{-iht} + |\varphi_1)e^{(-ih+\eta)t} + |\varphi_2)e^{(-ih+2\eta)t} + ...\;,
\end{equation}
where $|\varphi_j) = (h+ji\eta-H_0^{\rm SP})^{-1}W^{\rm SP}|\varphi_{j-1})$.
A state that starts in state $|X,h) e^{-iht}$ ends in a state $|\varphi)$ at $t=0$:
\begin{eqnarray}
|\varphi) &=& |X,h)
+ \int_{-\infty}^\infty \frac{dh'}{4\pi|h'|}\sum_{X'} \frac{(X',h'|W^{\rm SP}|X,h)}{h-h'+i\eta} |X',h')
\nonumber \\
&& + \int_{-\infty}^\infty\int_{-\infty}^\infty \frac{dh'\,dh''}{16\pi^2|h'h''|}\sum_{X'X''}
\frac{(X'',h''|W^{\rm SP}|X',h')(X',h'|W^{\rm SP}|X,h)}{(h-h''+2i\eta)(h-h'+i\eta)} |X'',h'')
+ ...\,.
\end{eqnarray}

Since in this problem we are interested in the {\em static} transmission probability at a given energy $h$, we must first take the limit as $\eta\rightarrow 0^+$ (that is, at $r_\star\rightarrow\pm\infty$, the perturbation term of Eq.~\ref{eq:C4} goes to zero, so the particle adiabatically enters and exits the potential). We emphasize that this is opposite from the order of limits in \citet{2024arXiv240709724V} where the physical limit was achieved by taking $\epsilon \rightarrow 0^+$ first. Setting $X={\rm up}$ and noting that the single-particle radial wave functions reduce to the limiting forms of Eqs.~(39) and (41) of \citet{PhysRevD.107.045004} in the $r_\ast \rightarrow \pm \infty$, the radial electron wave function in the $r_\ast \rightarrow \infty$ is given by
\begin{eqnarray}
\varphi
&\rightarrow& T_{\frac12kh} u_{\rm out}(h;r_\star)
+ \int_{-\infty}^\infty \frac{dh'}{4\pi|h'|}
\left[
\frac{({\rm up},h'|W^{\rm SP}|{\rm up},h)}{h-h'+i\eta} T_{\frac12kh'} u_{\rm out}(h';r_\star)
+
\frac{({\rm in},h'|W^{\rm SP}|{\rm up},h)}{h-h'+i\eta} R_{\frac12kh'} u_{\rm out}(h';r_\star)
\right]
\nonumber \\
&& + \int_{-\infty}^\infty\int_{-\infty}^\infty \frac{dh'\,dh''}{16\pi^2|h'h''|}\sum_{X'}\Bigl[
\frac{({\rm up},h''|W^{\rm SP}|X',h')(X',h'|W^{\rm SP}|{\rm up},h)}{(h-h''+2i\eta)(h-h'+i\eta)} 
T_{\frac12kh''} u_{\rm out}(h'';r_\star)
\nonumber \\ && ~~~~
+
\frac{({\rm in},h''|W^{\rm SP}|X',h')(X',h'|W^{\rm SP}|{\rm up},h)}{(h-h''+2i\eta)(h-h'+i\eta)} 
R_{\frac12kh''} u_{\rm out}(h'';r_\star)
\Bigr]
 + {\rm down},
\end{eqnarray}
where
\begin{equation}
u_{\rm out}(h;r_\star) = v^{-1/2} \left(\begin{array}c
\sqrt{h+\mu} \\ -i\sqrt{h-\mu} \end{array}\right) e^{i\zeta\ln(r_\star/2M)} e^{ipr_\star},
\end{equation}
with $p=\sqrt{h^2-\mu^2}$, $v=p/h$, and $\zeta=\mu^2M/p$ and the ``down'' terms behave as $e^{-ipr_\star}$. The changes to the behavior at large $r_\star$ (where the $e^{ipr_\star}$ term is rapidly oscillating) will come from the singularity when the final energy ($h'$ in the $1^{\text{st}}$-order term and $h''$ in the $2^{\text{nd}}$-order term) is near $h$. At this singularity, we may close the contour in the upper half complex plane (so that $e^{ipr_\star}\rightarrow 0$ as $\Im h>0$ and $\eta\rightarrow 0$), leading to
\begin{eqnarray}
\varphi
&\rightarrow& T_{\frac12kh} u_{\rm out}(h;r_\star)
\nonumber \\ &&
- i\frac{({\rm up},h+i\eta|W^{\rm SP}|{\rm up},h)}{2(h+i\eta)}
T_{\frac12k,h+i\eta} u_{\rm out}(h+i\eta;r_\star)
-i
\frac{({\rm in},h+i\eta|W^{\rm SP}|{\rm up},h)}{2(h+i\eta)}
R_{\frac12k,h+i\eta} u_{\rm out}(h+i\eta;r_\star)
\nonumber \\
&& - \int_{-\infty}^\infty \frac{i\,dh'}{8\pi|h'|(h+2i\eta)}\sum_{X'}\Bigl[
\frac{({\rm up},h+2i\eta|W^{\rm SP}|X',h')(X',h'|W^{\rm SP}|{\rm up},h)}{h-h'+i\eta} 
T_{\frac12k,h+2i\eta} u_{\rm out}(h+2i\eta;r_\star)
\nonumber \\ && ~~~~
+
\frac{({\rm in},h+2i\eta|W^{\rm SP}|X',h')(X',h'|W^{\rm SP}|{\rm up},h)}{h-h'+i\eta} 
R_{\frac12k,h+2i\eta} u_{\rm out}(h+2i\eta;r_\star)
\Bigr]
 + {\rm down}.
\end{eqnarray}
In this expression, formulae such as $(X',h+i\eta|W^{\rm SP}|X,h)$ are to be interpreted by solving for real $h'$ and {\em then} analytically continuing to $h'=h+i\eta$ (rather than the reverse, which does not lead to an analytic function of $h'$). Similarly, the analytic continuation of $|h'|$ in the vicinity of $+h$ is $h'$, not $|h'|$. Taking the limit as $\eta\rightarrow 0$, we find that the transmission coefficient has become, to second order,
\begin{eqnarray}
T^{\rm perturbed}(h) &=& T_{\frac12kh}
-i \frac{({\rm up},h|W^{\rm SP}|{\rm up},h)}{2h}
T_{\frac12kh} 
-i
\frac{({\rm in},h|W^{\rm SP}|{\rm up},h)}{2h}
R_{\frac12kh} 
\nonumber \\
&& - \int_{-\infty}^\infty \frac{i\,dh'}{8\pi|hh'|}\sum_{X'}\Bigl[
\frac{({\rm up},h|W^{\rm SP}|X',h')(X',h'|W^{\rm SP}|{\rm up},h)}{h-h'+i\eta} 
T_{\frac12kh}
\nonumber \\ &&
+ 
\frac{({\rm in},h|W^{\rm SP}|X',h')(X',h'|W^{\rm SP}|{\rm up},h)}{h-h'+i\eta}
R_{\frac12kh}\Bigr].
\end{eqnarray}
The transmission coefficient can be expanded in powers of $Z$ as:
\begin{equation}
{\mathbb T}_{\frac12kh}(Z) = 
{\mathbb T}_{\frac12kh}(0) + Z\partial_Z{\mathbb T}_{\frac12kh}(0) + \frac12 Z^2\partial_Z^2{\mathbb T}_{\frac12kh}(0) + ...\,,
\end{equation}
where the Taylor expansion terms are
\begin{align}
{\mathbb T}_{\frac12kh}(0) =& ~|T_{\frac12kh}|^2,
\nonumber \\
\partial_Z{\mathbb T}_{\frac12kh}(0) =& ~\frac1h \Im \Bigl[
R_{\frac12kh} T^\ast_{\frac12 kh}({\rm in},h|\underline W^{\rm SP}|{\rm up},h)
 \Bigr], ~~~~{\rm and}
\nonumber \\
\partial_Z^2{\mathbb T}_{\frac12kh}(0) =& ~\frac1h \Im
\int_{-\infty}^\infty \frac{dh'}{2\pi|h'|}\sum_{X'}\Bigl[
\frac{({\rm up},h|\underline W^{\rm SP}|X',h')(X',h'|\underline W^{\rm SP}|{\rm up},h)}{h-h'+i\eta} 
|T_{\frac12kh}|^2
\nonumber \\ &
~~~~+ 
\frac{({\rm in},h|\underline W^{\rm SP}|X',h')(X',h'|\underline W^{\rm SP}|{\rm up},h)}{h-h'+i\eta} R_{\frac12kh}T^\ast_{\frac12 kh}\Bigr]
\nonumber \\ &
+ \frac1{2h^2}\left|
T_{\frac12kh} 
({\rm up},h|\underline W^{\rm SP}|{\rm up},h)
+
R_{\frac12kh}({\rm in},h|\underline W^{\rm SP}|{\rm up},h)
 \right|^2.
\label{eq:Z-series}
\end{align}
(We have dropped terms that are manifestly real inside the imaginary part, and used an underline $\underline W^{\rm SP}$ to indicate that the $Z$ is not included.) We express the matrix elements $(Xh|\underline W^{\rm SP}|X'h')$ in terms of ${\cal I}^{(1)}_{XX'k}(h,h')$ and ${\cal I}^{(2)}_{XX'k}(h,h')$:
\begin{eqnarray}
(Xh|\underline W^{\rm SP}|X'h') &=& -\frac{\alpha}{2M}\sqrt{4hh'} {\cal I}^{(1)}_{XX'k}(h,h') 
\nonumber \\
&=& -\frac{\alpha}{2M}\sqrt{4hh'} \Bigl[
\frac{{\cal A}_{XX'k}(h,h')-i\epsilon \Lambda_{XX'k}(h,h)
-\frac12i\epsilon^2 N_{XX'k}(h)}{h-h'-i\epsilon}
+ 2\pi\delta_\epsilon(h-h') \delta_{X,\rm up}\delta_{X',\rm up} \Bigr]~~
{\rm and}
\nonumber \\ 
(Xh|\underline W^{\rm SP}|X',-h') &=& - \frac{\alpha}{2M}\sqrt{4hh'} {\cal I}^{(2)}_{XX'k}(h,h').
\label{eq:W_to_I}
\end{eqnarray}
Note the inclusion of the ${\bf N}$ term in Eq.~(\ref{eq:W_to_I}). As previously mentioned, we are taking the $\eta\rightarrow 0^+$ limit first unlike previous calculations (because we want to know what happens to the transmission coefficient at fixed energy, rather than allowing the change in potential at $r_\star \ll -2M$ to adiabatically raise or lower the energy of particles near the horizon). Therefore, when we take the limit as $\epsilon\rightarrow 0^+$, we are taking a limit as all 3 singularities ($h'=h+i\epsilon$, $h+i\eta$, and $h-i\epsilon$) approach each other, instead of just 2; and we keep terms in the numerator of Eq.~(\ref{eq:W_to_I}) through order $\epsilon^2$.

The $1^{\text{st}}$ derivative $\partial_Z {\mathbb T}_{\frac12 kh}(0)$ reduces to
\begin{equation}
\partial_Z{\mathbb T}_{\frac12kh}(0) = -\frac\alpha M \Im \Bigl[
R_{\frac12kh} T^\ast_{\frac12 kh} \Lambda_{{\rm in,up,}k}(h,h)
 \Bigr],
\label{eq:Z-der1}
\end{equation}
where the ${\cal A}$ term does not contribute due to the substitution of ${\cal A}_{{\rm in,up},k}(h,h)=i R^\ast_{\frac12kh} T_{\frac12kh}$ producing a term that does not have an imaginary component. Thus, we obtain an agreement with Eq.~(F20) of \citet{2024arXiv240709724V}.

\subsection{Simplifying the second derivative}
\label{app:2deriveT}

If the integrand in Eq.~(\ref{eq:Z-series}) is analytically continued, it has poles at $h' = h+i\epsilon$, $h+i\eta$, and $h-i\epsilon$. The equation for $\partial_Z^2{\mathbb T}_{\frac12kh}(0)$ in Eq.~(\ref{eq:Z-series}) can thus be broken into 3 terms:
\begin{equation}
\partial_Z^2{\mathbb T}_{\frac12kh}(0) =
\partial_Z^2{\mathbb T}_{\frac12kh}(0)\big|_{\rm int,NS} +
\partial_Z^2{\mathbb T}_{\frac12kh}(0)\big|_{\rm int,res} +
\partial_Z^2{\mathbb T}_{\frac12kh}(0)\big|_{\rm sqnorm},
\end{equation}
consisting of the integral with the non-singular term (contour deformed to pass below the pole at $h'=h-i\epsilon$); the residue contribution at $h'=h-i\epsilon$; and the square norm term. The square norm term simplifies straightforwardly:
\begin{eqnarray}
T_{\frac12kh} 
({\rm up},h|W^{\rm SP}|{\rm up},h)
+
R_{\frac12kh}({\rm in},h|W^{\rm SP}|{\rm up},h)
&=& -\frac{T_{\frac12kh}}{\epsilon} + T_{\frac12kh}\Lambda_{{\rm up,up},k}(h,h)+ R_{\frac12kh}\Lambda_{{\rm in,up},k}(h,h)
\nonumber \\
&&~~~~ + \frac\epsilon 2[T_{\frac12kh}N_{{\rm up,up},k}(h)+ R_{\frac12kh}N_{{\rm in,up},k}(h)],
\end{eqnarray} 
which leads to
\begin{eqnarray}
\partial_Z^2{\mathbb T}_{\frac12kh}(0)\Big|_{\rm sqnorm}
&=& \frac{\alpha^2}{M^2} \Biggl\{ 
\frac{|T_{\frac12kh}|^2}{2\epsilon^2}
+ \frac{|T_{\frac12kh}|^2\Lambda_{{\rm up,up},k}(h,h) + \Re \bigl[ R_{\frac12kh}T^\ast_{\frac12kh}\Lambda_{{\rm in,up},k}(h,h)\bigr]}{\epsilon}
\nonumber \\
&&~~~~ + \frac12
\bigl|T_{\frac12kh}\Lambda_{{\rm up,up},k}(h,h)
+R_{\frac12kh}\Lambda_{{\rm in,up},k}(h,h)\bigr|^2
\nonumber \\
&& ~~~~ + \frac12 |T_{\frac12kh}|^2N_{{\rm up,up},k}(h)+ \frac12 \Re \bigl[ R_{\frac12kh}T_{\frac12kh}^\ast N_{{\rm in,up},k}(h)\bigr] + {\cal O}(\epsilon)
\Biggr\}.
\end{eqnarray}
Using Eq.~(\ref{eq:w0}), we may re-write the ${\bf N}$ terms in terms of ${\bf Q}$ and the ${\boldsymbol\Lambda}$ terms in terms of the $2\times 2$ matrix ${\bf E}$ with entries ${\cal E}_{XX'k}(h)$:
\begin{eqnarray}
\partial_Z^2{\mathbb T}_{\frac12kh}(0)\Big|_{\rm sqnorm}
&=& \frac{\alpha^2}{M^2} \Biggl\{ 
\frac{|T_{\frac12kh}|^2}{2\epsilon^2}
+ \frac{|T_{\frac12kh}|^2\Lambda_{{\rm up,up},k}(h,h) + \Re \bigl[ R_{\frac12kh}T^\ast_{\frac12kh}\Lambda_{{\rm in,up},k}(h,h)\bigr]}{\epsilon}
\nonumber \\
&&~~~~ + \frac12
\bigl|T_{\frac12kh}{\cal E}_{{\rm up,up},k}(h)
+R_{\frac12kh}{\cal E}_{{\rm in,up},k}(h)\bigr|^2
\nonumber \\
&& ~~~~ - \frac12 \Im \bigl[ |T_{\frac12kh}|^2Q_{{\rm up,up},k}(h)+ R_{\frac12kh}T^\ast_{\frac12kh}Q_{{\rm in,up},k}(h)\bigr] + {\cal O}(\epsilon)
\Biggr\}.
\label{eq:Int-sqnorm}
\end{eqnarray}
The non-singular term is
\begin{eqnarray}
\partial_Z^2{\mathbb T}_{\frac12kh}(0)\Big|_{\rm int,NS} &=&
\frac{\alpha^2}{M^2} \Im \biggl[ \int_0^\infty \frac{dh'}{2\pi} \sum_{XX'} w_{Xkh}T^\ast_{\frac12kh} \frac{{\cal A}_{XX'k}(h,h') {\cal A}^\ast_{{\rm up},X'k}(h,h')}{(h-h'+i\upsilon)^3}
\nonumber \\
&& ~~~~~~~~ + \int_0^\infty \frac{dh'}{2\pi} \sum_{X'} R_{\frac12kh}T^\ast_{\frac12kh} \frac{{\cal I}^{(2)}_{{\rm in},X'k}(h,h') {\cal I}^{(2)\ast}_{{\rm up},X'k}(h,h')}{h+h'} \biggr],
\label{eq:Int-NS}
\end{eqnarray}
where we ignore the distinction between the poles at $h'=h+i\epsilon$ and $h'=h+i\eta$. As long as both poles are on the same side of the real axis, neither the distance from the real axis nor the order of the limits will alter the integral. 

For the residue contribution, we must carry out an analytic continuation. We define, for complex $h'$ (but still real $h$), $\bar{\cal A}_{XX'k}(h,h')$ to be the analytic continuation of the complex conjugate of ${\cal A}_{XX'k}(h,h')$. Note that this is different from the complex conjugate of the analytic continuation: the two are related by $\bar{\cal A}_{XX'k}(h,h') = [{\cal A}_{XX'k}(h,h'{^\ast})]^\ast$.
Then, the residue contribution is $-2\pi i$ times the residue (since the full integral minus the non-singular part is a clockwise path):
\begin{eqnarray}
    \partial_Z^2{\mathbb T}_{\frac12 kh}(0)\big\vert_{\rm int, res} &=& - \frac{\alpha^2}{M^2} \Re \mathop{\rm Res}_{h'=h-i\epsilon} \sum_{XX'}w_{Xkh}T^\ast_{\frac12 kh}
    \bigg[\frac{{\cal A}_{XX'k}(h,h')-i\epsilon\Lambda_{XX'k}(h,h) - \frac{i}{2}\epsilon^2 N_{XX'k}(h)}{h-h'-i\epsilon} \nonumber \\
    && ~~~ + 2\pi\delta_\epsilon(h-h')\delta_{X,\rm up}\delta_{X',\rm up}\Bigg]\Bigg[\frac{\bar{\cal A}_{{\rm up},X'k}(h,h')+i\epsilon\Lambda_{{\rm up},X'k}(h,h) + \frac{i}{2}\epsilon^2 N_{{\rm up},X'k}(h)}{h-h'+i\epsilon} \nonumber \\
    && ~~~ + 2\pi\delta_\epsilon(h-h')\delta_{X',\rm up}\bigg] \frac{1}{h-h'} ,
\end{eqnarray}
where we take the $\eta\rightarrow 0^+$ limit leading to the above expression. The residue is the coefficient of $1/z$ in the Laurent expansion around the pole, where $z = h-h'-i\epsilon$. We use the expansions:
\begin{equation}
2\pi\delta_\epsilon(h-h') = -\frac iz + \frac1{2\epsilon} + {\cal O}(z),
~~~
\frac1{h-h'} = -\frac i\epsilon + \frac z{\epsilon^2} + {\cal O}(z^2),
~~~{\rm and}~~~
\frac1{h-h'+i\epsilon} = -\frac i{2\epsilon}+{\cal O}(z) 
\end{equation}
to reduce this to
\begin{eqnarray}
\!\!\partial_Z^2{\mathbb T}_{\frac12 kh}(0)\Big\vert_{\rm int, res} &=& - \frac{\alpha^2}{M^2} \Re \mathop{\rm Res}_{h'=h-i\epsilon} \sum_{XX'}w_{Xkh}T^\ast_{\frac12 kh} \biggl[ \frac{{\cal A}_{XX'k}(h,h-i\epsilon)-i\delta_{X,\rm up}\delta_{X',\rm up} -i\epsilon\Lambda_{XX'k}(h,h) -\frac{i}2\epsilon^2 N_{XX'k}(h)}z \nonumber \\
&& ~~~ + \partial_{h'} {\cal A}_{XX'k}(h,h')\Big\vert_{h'=h-i\epsilon} + \frac{1}{2\epsilon}\delta_{X,\rm up}\delta_{X',\rm up}\biggr] \biggl[-\frac{i}z \delta_{X',\rm up}
-\frac i{2\epsilon}
{\cal A}^\ast_{{\rm up},X'k}(h,h+i\epsilon) \nonumber \\ && ~~~ +\frac12 \Lambda^\ast_{{\rm up},X'k}(h,h)+\frac\epsilon 2 N^\ast_{{\rm up},X'k}(h)
+ \frac{1}{2\epsilon}\delta_{X',\rm up} \biggr]\biggl[-\frac{i}{\epsilon} + \frac{z}{\epsilon^2}\biggr],
\end{eqnarray}
where the higher-order terms in $z$ do not contribute to the residue. Expanding the sum leads to
\begin{eqnarray}
    \partial_Z^2{\mathbb T}_{\frac12 kh}(0)\Big\vert_{\rm int, res} &=&  \frac{\alpha^2}{M^2} \Re\bigg\{\frac{1}{\epsilon^2} |T_{\frac12 kh}|^2 \left[i{\cal A}_{{\rm up,up},k}(h,h-i\epsilon)+1 +\epsilon\Lambda_{{\rm up,up},k}(h,h) + \frac12\epsilon^2 N_{{\rm up,up},k}(h)\right] \nonumber \\
    && + \frac{1}{\epsilon^2} R_{\frac12 kh}T^\ast_{\frac12 kh} \left[i{\cal A}_{{\rm in,up},k}(h,h-i\epsilon) +\epsilon\Lambda_{{\rm in,up},k}(h,h)
+\frac12\epsilon^2 N_{{\rm in,up},k}(h)
\right] \nonumber \\
&& + \frac{i}{2\epsilon} |T_{\frac12 kh}|^2 \left[{\cal A}_{{\rm up,up},k}(h,h-i\epsilon)-i -i\epsilon\Lambda_{{\rm up,up},k}(h,h)
-\frac12i\epsilon^2 N_{{\rm up,up},k}(h)\right] \nonumber \\
&& ~~~~~~~~\times
\left[\frac {-i{\cal A}^\ast_{{\rm up,up},k}(h,h+i\epsilon)+1 }{\epsilon} + \Lambda^\ast_{{\rm up,up},k}(h,h)
+ \frac12\epsilon N^\ast_{{\rm up,up},k}(h) \right] \nonumber \\
&& + \frac{i}{2\epsilon} |T_{\frac12 kh}|^2 \left[{\cal A}_{{\rm up,in},k}(h,h-i\epsilon)-i\epsilon\Lambda_{{\rm up,in},k}(h,h)
-\frac12i\epsilon^2 N_{{\rm up,in},k}(h)\right] \nonumber \\
&& ~~~~~~~~\times
\left[\frac {-i{\cal A}^\ast_{{\rm up,in},k}(h,h+i\epsilon)}{\epsilon} + \Lambda^\ast_{{\rm up,in},k}(h,h)
+ \frac12\epsilon N^\ast_{{\rm up,in},k}(h) \right] \nonumber \\
&& + \frac{i}{2\epsilon} R_{\frac12 kh}T^\ast_{\frac12 kh} \left[{\cal A}_{{\rm in,up},k}(h,h-i\epsilon) -i\epsilon\Lambda_{{\rm in,up},k}(h,h)  -\frac12i\epsilon^2 N_{{\rm in,up},k}(h)
\right] \nonumber \\
&& ~~~~~~~~\times
\left[\frac {-i{\cal A}^\ast_{{\rm up,up},k}(h,h+i\epsilon)+1 }{\epsilon} + \Lambda^\ast_{{\rm up,up},k}(h,h)
+ \frac12\epsilon N^\ast_{{\rm up,up},k}(h) \right] \nonumber \\
&& + \frac{i}{2\epsilon}R_{\frac12 kh}T^\ast_{\frac12 kh}\left[{\cal A}_{{\rm in,in},k}(h,h-i\epsilon) -i\epsilon\Lambda_{{\rm in,in},k}(h,h)
-\frac12i\epsilon^2 N_{{\rm in,in},k}(h)
\right] \nonumber \\
&& ~~~~~~~~\times
\left[\frac {-i{\cal A}^\ast_{{\rm up,in},k}(h,h+i\epsilon) }{\epsilon} + \Lambda^\ast_{{\rm up,in},k}(h,h)
+ \frac12\epsilon N^\ast_{{\rm up,in},k}(h) \right] \nonumber \\
&& +\frac{1}{\epsilon}|T_{\frac12kh}|^2
\left[
\partial_{h'}{\cal A}_{{\rm up,up},k}(h,h')\Big|_{h'=h-i\epsilon}
+ \frac1{2\epsilon}
\right]
 +\frac{1}{\epsilon}R_{\frac12kh}T_{\frac12kh}^\ast \partial_{h'}{\cal A}_{{\rm in,up},k}(h,h')\Big|_{h'=h-i\epsilon}
\bigg\}.~~~
\label{eq:DZ2}
\end{eqnarray}
This can be expanded in powers of $\epsilon$, using the Taylor series for ${\cal A}_{XX'k}(h,h')$ around $h'=h$. Expressing the collection of $\cal A$ terms as a $2\times 2$ matrix (with ``in'' first, then ``up''), we have
\begin{equation}
\begin{split}
    &{\bf A}_k(h,h\pm i\epsilon) = {\bf A}^{(0)}_k(h) \pm i \epsilon [{\bf \Upsilon}_k(h)-{\bf \Lambda}_k(h,h)]-\frac12 \epsilon^2 {\bf Q}_k(h) + \mathcal{O}(\epsilon^3) ~~{\rm and} \\
    \ & \partial_{h'}{\bf A}_k(h,h')\Big\vert_{h'=h\pm i\epsilon} = {\bf \Upsilon}_k(h) - {\bf \Lambda}_k(h,h) \pm i\epsilon {\bf Q}_k(h) + \mathcal{O}(\epsilon^2).
\end{split}
\end{equation}
The expansion coefficients for terms that are $0^{\text{th}}$ order in $\epsilon$ come from Eq.~(\ref{eq:A-val}):
\begin{equation}
{\bf A}^{(0)}_k(h) = \left(
\begin{array}{cc}
-i|T_{\frac12kh}|^2 & iR^\ast_{\frac12kh}T_{\frac12kh}
\\
iT^\ast_{\frac12kh}R_{\frac12kh} & i|T_{\frac12kh}|^2
\end{array}
\right).
\end{equation}
Terms that are $1^{\text{st}}$ order in $\epsilon$ come from Eq.~(\ref{eq:E2}):
\begin{eqnarray}
\boldsymbol \Upsilon_k(h)
= {\bf E}_k(h) + \boldsymbol\Lambda_k(h,h)
&=& 
 \left(\begin{array}{cc} |T_{\frac12kh}|^2 & -2R^\ast_{\frac12kh}T_{\frac12kh} \\
-T^\ast_{\frac12kh}R_{\frac12kh} & 2|R_{\frac12kh}|^2
\end{array} \right) \frac{d}{dh}\arg T_{\frac12kh}
+ \left(\begin{array}{cc} 0 & R^\ast_{\frac12kh}T_{\frac12kh} \\
0 & -|R_{\frac12kh}|^2
\end{array} \right) \frac{d}{dh}\arg R_{\frac12kh}
\nonumber \\ && ~~+
\frac i2 \left(\begin{array}{cc}
-1 & -T_{\frac12kh}/R_{\frac12kh} \\
R_{\frac12kh}/T_{\frac12kh} & 1
\end{array}\right)
\frac{d{\mathbb T}_{\frac12kh}}{dh}.
\end{eqnarray}
Substituting these values into Eq.~(\ref{eq:DZ2}) provides a mass cancellation, particularly using the lemma of Eq.~(\ref{eq:w0}) to cancel out some of the combinations of $\boldsymbol\Upsilon$ and ${\bf N}-i{\bf Q}$,
resulting in:
\begin{eqnarray}
\partial_Z^2{\mathbb T}_{\frac12kh}(0)\Big|_{\rm int,res} &=&
\frac{\alpha^2}{M^2} \biggl\{
-\frac{|T_{\frac12kh}|^2}{2\epsilon^2}
- \frac{
|T_{\frac12kh}|^2 \Lambda_{{\rm up,up},k}(h,h)
+ \Re \bigl[ R_{\frac12kh}T^\ast_{\frac12kh}\Lambda_{{\rm in,up},k}(h,h) \bigr]
}\epsilon 
\nonumber \\
&& +
\Im \bigl[  
|T_{\frac12 kh}|^2 Q_{{\rm up,up,}k}(h)
+ R_{\frac12kh}T^\ast_{\frac12 kh} Q_{{\rm in,up,}k}(h) \bigr]
+ {\cal O}(\epsilon)  \biggr\} .
\label{eq:Int-res}
\end{eqnarray}
Finally, by combining Eqs.~(\ref{eq:Int-sqnorm}), (\ref{eq:Int-NS}), and (\ref{eq:Int-res}), we arrive at the final result:
\begin{eqnarray}
\partial_Z^2{\mathbb T}_{\frac12kh}(0)
&=& \frac{\alpha^2}{M^2} \biggl\{ 
 \frac12
\bigl|T_{\frac12kh}{\cal E}_{{\rm up,up},k}(h,h)
+R_{\frac12kh}{\cal E}_{{\rm in,up},k}(h,h)\bigr|^2
 + \frac12 \Im \left[ |T_{\frac12kh}|^2Q_{{\rm up,up},k}(h)+  R_{\frac12kh}T_{\frac12kh}^\ast Q_{{\rm in,up},k}(h)\right] 
\nonumber \\
&& +
\Im \biggl[ \int_0^\infty \frac{dh'}{2\pi} \sum_{XX'} w_{Xkh}T^\ast_{\frac12kh} \frac{{\cal A}_{XX'k}(h,h') {\cal A}^\ast_{{\rm up},X'k}(h,h')}{(h-h'+i\upsilon)^3}
\nonumber \\
&& ~~~~~~~~ + \int_0^\infty \frac{dh'}{2\pi} \sum_{X'} R_{\frac12kh}T^\ast_{\frac12kh}  \frac{{\cal I}^{(2)}_{{\rm in},X'k}(h,h') {\cal I}^{(2)\ast}_{{\rm up},X'k}(h,h')}{h+h'} \biggr]
\biggr\}
\label{eq:Z2-der}
\end{eqnarray}
which describes how the semi-classical derivative $\partial_Z^2{\mathbb T}_{\frac12 kh}(0)$ relates to quantities that appear in our QFT approach.

\section{Effect of a $1/r^2$ potential on the transmission probability}
\label{app:1r2}

In this section, we investigate how the transmission probability ${\mathbb T}_{\frac12kh}$ changes due to the perturbation
\begin{equation}
W^{\rm IS} = -\frac{2M \alpha Y}{r^2} e^{-2\epsilon |r_\star|},
~~~~ \underline W^{\rm IS} = -\frac{2M}\alpha (\underline W^{\rm SP})^2,
\label{eq:WIS}
\end{equation}
where $Y$ is the amplitude of the perturbation. Physically, a $\sim 1/r^2$ perturbation corresponds to the classical self-energy of the electron and can be written in terms of the square of the Coulomb potential, Eq.~(\ref{eq:C4}). This modified potential appears when simplifying $B_0(k,h)$ in \S\ref{sec:B0_derivation} and must be considered when expressing our full QFT result in terms of semi-classical derivatives.
The change in transmission probability follows the same methodology as in Eq.~(\ref{eq:Z-series}) except that we consider the matrix element of $\underline{W}^{IS}$ instead of $\underline{W}^{\rm SP}$:
\begin{equation}
\partial_Y {\mathbb T}_{\frac12kh} = \frac1h \Im \Bigl[ R_{\frac12kh}T_{\frac12kh}^\ast  ({\rm in},h| \underline W^{\rm IS} | {\rm up},h) \Bigr].
\end{equation}
This can be expanded using the insertion of the identity operator,
\begin{equation}
\underline W^{\rm IS} = -\frac{2M}\alpha \sum_{X'} \int_{-\infty}^\infty \frac{dh'}{4\pi |h'|} \,\underline W^{\rm SP} |X',h')(X',h'| \underline W^{\rm SP},
\end{equation}
which leads to
\begin{eqnarray}
\partial_Y {\mathbb T}_{\frac12kh} &=& -\frac\alpha{M} \Im \Bigl\{ R_{\frac12kh}T_{\frac12kh}^\ast  \sum_{X'}
\int_0^\infty \frac{dh'}{2\pi}\,\frac{{\cal A}_{{\rm in},X',k}(h,h') - i\epsilon\Lambda_{{\rm in},X'k}(h,h)}{h-h'-i\epsilon}
\nonumber \\ && ~~~~~~~~\times
\Bigl[ \frac{{\cal A}_{{\rm up},X',k}(h,h') - i\epsilon\Lambda_{{\rm up},X'k}(h,h)}{h-h'-i\epsilon} + 2\pi\delta_\epsilon(h-h') \delta_{X',\rm up}
\Bigr]^\ast
\nonumber \\
&& + R_{\frac12kh}T_{\frac12kh}^\ast  \sum_{X'} \int_0^\infty \frac{dh'}{2\pi}\,{\cal I}^{(2)}_{{\rm in},X',k}(h,h') {\cal I}^{(2)\ast}_{{\rm up},X',k}(h,h') 
 \Bigr\}.
\end{eqnarray}
This integral can be expanded into 6 types of terms depending on the powers of each quantity: ${\cal A}^2$, $\Lambda{\cal A}$, $\Lambda^2$, ${\cal A}\delta$, $\Lambda\delta$, and ${\cal I}^{(2)\,2}$. 
The power-counting in $\epsilon$ shows that the $\Lambda^2$ terms trivially drop out. Writing the remaining terms in this order, we obtain:
\begin{eqnarray}
\partial_Y {\mathbb T}_{\frac12kh} &=& -\frac\alpha{M} \Im \Bigl\{ R_{\frac12kh}T_{\frac12kh}^\ast  \sum_{X'}
\int_0^\infty \frac{dh'}{2\pi}\, \frac{
{\cal A}_{{\rm in},X',k}(h,h'){\cal A}_{{\rm up},X',k}^\ast(h,h')
}{(h-h'-i\epsilon)(h-h'+i\epsilon)}
\nonumber \\
&& + i\epsilon R_{\frac12kh}T_{\frac12kh}^\ast \sum_{X'} 
\int_0^\infty \frac{dh'}{2\pi}\, \frac{
-\Lambda_{{\rm in},X'k}(h,h) 
{\cal A}_{{\rm up},X',k}^\ast(h,h')
+ \Lambda_{{\rm up},X'k}^\ast(h,h) {\cal A}_{{\rm in},X',k}(h,h')
}{(h-h'-i\epsilon)(h-h'+i\epsilon)}
\nonumber \\
&& + R_{\frac12kh}T_{\frac12kh}^\ast 
\int_0^\infty dh'\, \frac{\delta_\epsilon(h-h')}{h-h'-i\epsilon} {\cal A}_{{\rm in,up},k}(h,h')
 -i\epsilon R_{\frac12kh}T_{\frac12kh}^\ast  \Lambda_{{\rm in,up},k}(h,h)
\int_0^\infty dh'\, \frac{\delta_\epsilon(h-h')}{h-h'-i\epsilon}
\nonumber \\
&& + R_{\frac12kh}T_{\frac12kh}^\ast  \sum_{X'} \int_0^\infty \frac{dh'}{2\pi}\,{\cal I}^{(2)}_{{\rm in},X',k}(h,h') {\cal I}^{(2)\ast}_{{\rm up},X',k}(h,h') 
 \Bigr\}.
 \label{eq:temp-x1}
\end{eqnarray}
The first two terms may be simplified using Eq.~(\ref{eq:PPI}), and the specific values of the numerators when $h'=h$ (which cancel out when the imaginary part is taken). The third and fourth terms may be simplfied using Eq.~(\ref{eq:-i}), leaving
\begin{eqnarray}
\partial_Y {\mathbb T}_{\frac12kh} &=& -\frac\alpha{M} \Im \Bigl\{ R_{\frac12kh}T_{\frac12kh}^\ast  \sum_{X'}
{\mathbb P}\int_0^\infty \frac{dh'}{2\pi}\, \frac{
{\cal A}_{{\rm in},X',k}(h,h'){\cal A}_{{\rm up},X',k}^\ast(h,h')
}{(h-h')^2}
\nonumber \\
&& ~~+ R_{\frac12kh}T_{\frac12kh}^\ast \Lambda_{{\rm in,up},k}(h,h) + \frac{i}{4}{\mathbb T}_{\frac12 k h} \frac{d {\mathbb T}_{\frac12 kh}}{dh}
+ R_{\frac12kh}T_{\frac12kh}^\ast  \sum_{X'} \int_0^\infty \frac{dh'}{2\pi}\,{\cal I}^{(2)}_{{\rm in},X'k}(h,h') {\cal I}^{(2)\ast}_{{\rm up},X'k}(h,h') 
 \Bigr\},~~~
\label{eq:dyt-final}
\end{eqnarray} 
where we have dropped terms that do not contribute to the imaginary part. Half of the $R_{\frac12kh}T_{\frac12kh}^\ast \Lambda_{{\rm in,up},k}(h,h)$ term and the $d{\mathbb T}_{\frac12kh}/dh$ term come from the ${\cal A}\delta$ term in Eq.~(\ref{eq:temp-x1}) while the other half of the $R_{\frac12kh}T_{\frac12kh}^\ast \Lambda_{{\rm in,up},k}(h,h)$ term comes from the $\Lambda\delta$ term. Using Eq.~(\ref{eq:Z-der1}), it is perhaps more illuminating to write our derived expression as
\begin{eqnarray}
\partial_Y {\mathbb T}_{\frac12kh} &=& - \frac{\alpha}{4M} {\mathbb T}_{\frac12 kh} \frac{d{\mathbb T}_{\frac12 kh}}{dh} + \partial_Z {\mathbb T}_{\frac12 kh}(0)
\nonumber \\
&& -\frac{\alpha}{M} \Im \Bigr\{ R_{\frac12kh}T_{\frac12kh}^\ast  \sum_{X'} {\mathbb P} \int_0^\infty \frac{dh'}{2\pi}\, \left[{\cal I}^{(1)}_{{\rm in},X'k}(h,h') {\cal I}^{(1)\ast}_{{\rm up},X'k}(h,h') + {\cal I}^{(2)}_{{\rm in},X'k}(h,h') {\cal I}^{(2)\ast}_{{\rm up},X'k}(h,h') \right] 
 \Bigr\}.
\label{eq:dyt-final2}
\end{eqnarray} 
Note that we are only interested in the principal part of the total integral and are allowed to express ${\cal A}_{XX'}(h,h') = {\cal I}^{(1)}_{XX'}(h,h')/(h-h')$ as we are neglecting any potential singularities that might appear at $h=h'$.

\section{Semi-classical terms}
\label{sec:sc_terms}

\citet{2024arXiv240709724V} \S4 presented the perturbative description of the \citet{PhysRevD.16.2402} semiclassical electron spectrum. Here we present the result in terms of derivatives of ${\mathbb T}_{\frac12 kh}(Z)$ and $f_{\rm up}(h)$. The correction derived in \citet{2024arXiv240709724V} is
%
\begin{equation}
 \frac{dN_{e^\pm}}{dhdt}\bigg|_{\rm sc} = \alpha \left(\frac{dN_{[1]} }{dhdt} + \alpha \langle Z^2 \rangle \frac{dN_{[2]}}{dhdt} \right) + \mathcal{O}(\alpha^2),
\end{equation}
where 
\begin{equation}
    \frac{dN_{[\lambda]}}{dhdt} = \frac{1}{2\pi}\frac1{ \lambda!} \frac{\partial^\lambda}{\partial (\alpha Z)'^\lambda} \sum_k (2j+1) f_{\rm up}\left(h + \frac{\alpha Z'}{2M}\right) \bigg\vert_{Z'=0}
\end{equation}
and $f_{\rm up}(h) = 1/(e^{8\pi M h} +1)$.
Expanding the derivatives, we obtain
\begin{eqnarray}
    \frac{dN_{e^\pm}}{dhdt}\bigg|_{\rm sc} &=& \frac{1}{2\pi} \sum_k (2j+1) \bigg\{[\partial_{Z} {\mathbb T_{\frac12 kh}}(0)]f_{\rm up}(h) + \frac{\alpha}{2M} {\mathbb T}_{\frac12 kh}(0) \frac{df_{\rm up}}{dh}  \nonumber \\
    && + \langle Z^2 \rangle \bigg[ \frac12 [\partial^2_{Z} {\mathbb T}_{\frac12 kh}(0)]f_{\rm up}(h) + \frac{\alpha}{2M} [\partial_{Z} {\mathbb T}_{\frac12 kh}(0)] \frac{df_{\rm up}}{dh} + \frac{\alpha^2}{8M^2} {\mathbb T}_{\frac12 kh}(0) \frac{d^2 f_{\rm up}}{dh^2} \bigg] \bigg\} + {\cal O}(\alpha^2).
\label{eq:sc_spectra}
\end{eqnarray}
Since $\partial_Z^\lambda {\mathbb T}_{\frac12 kh}$ is ${\cal O}(\alpha^\lambda)$ and $\langle Z^2 \rangle \sim \mathcal{O}(\alpha^{-1})$, the terms written explicitly in $dN_{\rm sc}/dhdt$ are $\mathcal{O}(\alpha)$.

\section{Truncation errors for $r$ and $k$}
\label{sec:rk_error}

Here we consider the errors from setting our outer boundary condition at $r_{\rm max}$ and using a maximum value of $|k|$. To assess the impact of $r_{\rm max}$, we note that the constant probability current $\psi^2 v$, means that in the non-relativistic limit ($h-\mu \ll \mu$, $2M/r\ll 1$):
\begin{equation}
    \psi \propto \frac{1}{v^{1/2}} \propto \bigl(h-\mu+\frac{M\mu}{r}\bigr)^{-1/4}.
\end{equation}
This means that the ratio of the amplitude of $\psi$ and its asymptotic amplitude in this limit is
\begin{equation}
    \frac{\text{Amplitude}}{\text{Asymptotic Amplitude}}=\frac{(h-\mu+\frac{M\mu}{r})^{-1/4}}{(h-\mu)^{-1/4}}
    ~\xrightarrow{{\rm large}~r}~
    1-\frac{\mu M}{4r(h-\mu)}+...
\end{equation}
and therefore to get a fractional error in amplitude $\epsilon_\text{err}$, we need
\begin{equation}
    r_{\text{max}}>\frac{\mu M}{4\epsilon_\text{err}(h-\mu)}.
\end{equation}
The $r_\text{max}$ value of $10^6M$ means that our fractional error is less than $0.035\%$ at our lowest positive value of $h-\mu$. It is therefore sufficiently high to mitigate the outer boundary effects in the non-relativistic limit outlined above. 

Terms with higher values of $|k|$ contribute less to the $\mathcal{O}(\alpha)$ correction to the spectrum than terms with lower values of $|k|$. The terms with $|k|=5$ have a maximum relative contribution of 0.0058\% in the $M=8\times10^{21}m_P$ case and 0.0013\% in the $M=1\times10^{21}m_P$ case. Terms with higher values of $|k|$ would therefore have negligible contribution to the spectrum correction, and it is sufficient to include only terms with $|k|\leq5$.

\bibliography{main}

\end{document}